\documentclass[12pt]{article}

\newcommand{\be}{\begin{equation}}
\newcommand{\ee}{\end{equation}}
\newcommand{\Dlt}{\Delta}
\newcommand{\dlt}{\delta}

\newcommand{\vp}{\varphi}
\newcommand{\ep}{\varepsilon}
\newcommand{\al}{\alpha}
\newcommand{\ra}{\rightarrow}

\newcommand{\lbd}{\lambda}
\newcommand{\Lbd}{\Lambda}
\newcommand{\cD}{{\cal D}}
\newcommand{\cC}{{\cal C}}

\newcommand{\cL}{{\cal L}}
\newcommand{\cA}{{\cal A}}
\newcommand{\cM}{{\cal M}}
\newcommand{\rgl}{\rangle}
\newcommand{\lgl}{\langle}

\usepackage{amssymb}
\usepackage{amsmath}
\usepackage{amscd}
\usepackage{latexsym}
\usepackage{cite}

\begin{document}

\begin{center}

{\Large{\bf
Processing Information in Quantum Decision Theory} \\ [5mm]

Vyacheslav I. Yukalov $^{1,2}$ and Didier Sornette $^{1,3}$} \\ [3mm]

{\it
$^{1}$ Department of Management, Technology and Economics \\
ETH Z\"urich, Z\"urich CH-8032, Switzerland \\
E-Mail: syukalov@ethz.ch \\ [3mm]

$^{2}$ Bogolubov Laboratory of Theoretical Physics \\
Joint Institute for Nuclear Research, Dubna 141980, Russia \\
E-Mail: yukalov@theor.jinr.ru \\ [3mm]

$^{3}$ Swiss Finance Institute, c/o University of Geneva \\
 40 blvd. Du Pont d'Arve, CH 1211 Geneva 4, Switzerland}

\end{center}

\vskip 2cm

\begin{abstract}

A survey is given summarizing the state of the art of
describing information processing in Quantum Decision Theory, which
has been recently advanced as a novel variant of decision making,
based on the mathematical theory of separable Hilbert spaces. This
mathematical structure captures the effect of superposition of
composite prospects, including many incorporated intended actions.
The theory characterizes entangled decision making,
non-commutativity of subsequent decisions, and intention
interference. The self-consistent procedure of decision making, in
the frame of the quantum decision theory, takes into account both
the available objective information as well as subjective contextual
effects. This quantum approach avoids any paradox typical of
classical decision theory. Conditional maximization of entropy,
equivalent to the minimization of an information functional, makes
it possible to connect the quantum and classical decision theories,
showing that the latter is the limit of the former under vanishing
interference terms.
\end{abstract}

\vskip 1cm

{\parindent=0pt
{\bf Keywords}: quantum information processing; quantum decision making;
entangled decisions; intention interference; decision
non-commutativity; minimal information

\vskip 1cm

}

\newpage

\section{Basic Ideas and Historical Retrospective}

This section serves as an introduction to the problem, describing the
historical retrospective, related studies, and basic ideas of the
quantum approach to decision making. First of all, in order that the
reader would not be lost in details, we need to stress the main goal
of the approach.

\vskip 2mm

{\bf Principal Goal:} {\it The principal goal of the quantum approach
to decision making is to develop a unified theory that, from one side,
could formalize the process of taking decisions by human decision makers
in terms of quantum language and, from another side, would suggest a
scheme of thinking quantum systems that could be employed for creating
artificial intelligence}.

\vskip 2mm

Generally, decision theory is concerned with identifying what are
the optimal decisions and how to reach them. Traditionally, it is a
part of discrete mathematics. Most of decision theory is normative
and prescriptive, and assumes that people are fully-informed and
rational. These assumptions have been questioned early on with the
evidence provided by the Allais paradox \cite{1} and many other
behavioral paradoxes \cite{2}, showing that humans often seem to
deviate from the prescription of rational decision theory due to
cognitive and emotion biases. The theories of bounded rationality
\cite{3}, of behavioral economics and of behavioral finance have
attempted to account for these deviations. As reviewed by Machina
\cite{4}, alternative models of preferences over objectively or
subjectively uncertain prospects have attempted to accommodate these
systematic departures from the expected utility model, while
retaining as much of its analytical power as possible. In
particular, non-additive nonlinear probability models have been
developed to account for the deviations from objective to subjective
probabilities observed in human agents \cite{5,6,7,8,9,10}. However,
many paradoxes remain unexplained or are sometimes rationalized on
an ad hoc basis, which does not provide much predictive power.

Another approach to decision theory can be proposed, being part of
the mathematical theory of Hilbert spaces \cite{11} and employing
the mathematical techniques that are used in quantum theory. (see,
e.g., the special issue \cite{11bis} and references therein).
However, no self-consistent quantum theory of decision making has
been developed, which would have predictive power.

Recently, we introduced a general framework, called the
{\it Quantum Decision Theory} (QDT), in which decisions involve composite
intended actions which, as we explain below, provides a unifying explanation
of many paradoxes of classical decision theory in a quantitative predictive
manner \cite{12}. Such an approach can be thought of as the mathematically
simplest and most natural extension of objective probabilities into nonlinear
subjective probabilities. The proposed formalism allows one to explain
{\it quantitatively} different anomalous phenomena, e.g., the disjunction
and conjunction effects. The disjunction effect is the failure of humans to
obey the sure-thing principle of classical probability theory. The conjunction
effect is a logical fallacy that occurs when people assume that specific
conditions are more probable than a single general one. The QDT
unearths a deep relationship between the conjunction fallacy and the
disjunction effect, the former being sufficient for the latter to exist.

QDT uses the same underlying mathematical structure as the one
developed to establish a rigorous formulation of quantum mechanics
\cite{13}. Based on the mathematical theory of separable Hilbert
spaces on the continuous field of complex numbers, quantum mechanics
showed how to reconcile and combine the continuous wave description
with the fact that waves are organized in discrete energy packets,
called quanta, which behave in a manner similar to particles.
Analogously, in the QDT framework, the qualifier quantum emphasizes
the fact that a decision is a discrete selection from a large set of
entangled options. The key idea of QDT is to provide the simplest
generalization of the classical probability theory underlying
decision theory, so as to account for the complex dynamics of the
many nonlocal hidden variables that may be involved in the cognitive
and decision making processes of the brain. The mathematical theory
of complex separable Hilbert spaces provides the simplest direct way
to avoid dealing with the unknown hidden variables, and at the same
time reflecting the complexity of nature \cite{14}. In
decision making, unknown states of nature, emotions, and
subconscious processes play the role of hidden variables.

Before presenting the QDT approach, it is useful to briefly summarize
previous studies of decision making and of the associated cognitive
processes of the brain which, superficially, could be considered as
related to the QDT approach. This exposition will allow us to underline
the originality and uniqueness of the approach. We do not touch here purely
physiological aspects of the problem, which are studied in medicine and
physiological cognitive sciences. Concerning the functional aspects of the
brain, we focus our efforts towards its formal mathematical modeling.

One class of approaches is based on the theory of neural networks
and of dynamical systems (see, e.g., \cite{15,16,17,18}).
These bottom-up approaches suffer from the obvious difficulties of
modeling the emergence of upper mental faculties from a microscopic
constructive neuron-based description.

Two main classes of theories invoke the qualifier ``quantum''. In
the first class, one finds investigations which attempt to represent
the brain as a quantum or quantum-like object \cite{19,20,21}, for
which several mechanisms have been suggested
\cite{22,23,24,25,26,27,28}. The existence of genuine quantum
effects and the operation of any of these mechanisms in the brain
remain however controversial and have been criticized by Tegmark as
being unrealistic \cite{29}. Another approach in this first class
appeals to the mind-matter duality, treating mind and matter as
complementary aspects and considering consciousness as a separate
fundamental entity \cite{30,31,32,33}. This allows one, without
insisting on the quantum nature of the brain processes, if any, to
ascribe quantum properties solely to the consciousness itself, as
has been advocated by Stapp \cite{34,35}. Actually, the basic idea
that mental processes are similar to quantum-mechanical phenomena
goes back to the founder of the old quantum mechanics, Niels Bohr.
One of the first publications on this analogy is his paper
\cite{36}. Later on, he returned many times to the similarity
between quantum mechanics and the function of the brain, for
instance in \cite{37,38,39}. This analogy proposes that mental
processes could be modeled by a quantum-mechanical wave function,
whose evolution would be characterized by a dynamical equation, like
the Schr\"odinger equation.

The second class of theories does not necessarily assume quantum
properties of the brain or that consciousness is a separate entity
with quantum characteristics. Rather, these approaches use quantum
techniques, as a convenient language to generalize classical
probability theory. An example is provided by the quantum games
\cite{40,41,42,43,44,45,46,47,48,49}. With the development of
quantum game theory, it has been shown that many quantum games can
be reformulated as classical games by allowing for a more complex
game structure \cite{50,50b,50c,50d}. But, in the majority of cases,
it is more efficient to play quantum game versions, as less
information needs to be exchanged. Another example is the Shor
algorithm \cite{51}, which is purely quantum-mechanical but is
solving the classical factoring problem. This shows that there is no
contradiction in using quantum techniques to describe classical
problems. Here ``classical'' is contrasted with ``quantum'', in the
sense consecrated by decades of discussions on the interpretation of
quantum mechanics. In fact, some people go as far as stating that
quantum mechanics is nothing but an effective theory describing very
complicated classical systems \cite{52,53,54}. Interpretations of
this type have been made, e.g., by de Broglie and Bohm. An extensive
literature in this direction can be found in de Broglie
\cite{55} and Bohm \cite{56}. In any case, whether we deal really
with a genuinely quantum system or with an extremely complex
classical system, the language of quantum theory can be a convenient
effective tool for describing such systems \cite{14}. In the case of
decision making performed by real people, the subconscious activity
and the underlying emotions, which are difficult to quantify, play
the role of the hidden variables appearing in quantum theory.

The QDT belongs to this second class of theories, {\it i.e.}, we use
the construction of complex separable Hilbert spaces as a {\it
mathematical language} that is convenient for characterizing the
complicated processes in the mind, which are associated with
decision making. This approach encompasses in a natural way several
delicate features of decision making, such as its probabilistic
nature, the existence of entangled decisions, the possible
non-commutativity of decisions, and the interference between several
different decisions. These terms and associated concepts are made
operationally clear in the sequel.

As a bonus, the QDT provides natural algorithms which could be
used in the future for quantum information processing, the operation of
quantum computers, and in creating artificial intelligence.

The classical approaches to decision making are based on utility
theory \cite{57,58}. Decision making in the presence of uncertainty
about the states of nature is formalized in the statistical decision
theory \cite{59,60,61,62,63,64,65,66,67,68,69}. Some
problems, occurring in the interpretation of the classical utility
theory and its application to real human decision
processes have been discussed in numerous 
literature~(e.g., \cite{64,70,71,4}).

Quantum approach to decision making, suggested in Reference
\cite{12}, is principally different from the classical utility
theory. In this approach, the action probability is defined as is
done in quantum mechanics, using the mathematical theory of complex
separable Hilbert spaces. This proposition can be justified by
invoking the following analogy. The probabilistic features of
quantum theory can be interpreted as being due to the existence of
the so-called nonlocal hidden variables. The dynamical laws of these
nonlocal hidden variables could be not merely extremely cumbersome,
but even not known at all, similarly to the unspecified states of
nature in decision theory. The formalism of quantum theory is then
formulated in such a way as to avoid dealing with unknown hidden
variables, but at the same time, to reflect the complexity of nature
\cite{14}. In decision making, the role of hidden variables is
played by unknown states of nature, by emotions, and by subconscious
processes, for which quantitative measures are not readily
available.

In the following sections, we develop the detailed description of
the suggested program, explicitly constructing the action
probability in quantum-mechanical terms. The probability of an
action is intrinsically subjective, as it must characterize intended
actions by human beings. For brevity, an intended action can be
called an {\it intention} or just an {\it action}. And, in
compliance with the terminology used in the theories of
decision-making, a composite set of intended actions, consisting of
several sub-actions, is called a {\it prospect}. An important
feature of the quantum approach is that, in general, it deals not
with separate intended actions, but with composite prospects,
including many incorporated intentions. Only then it becomes
possible, within the frame of one general theory, to describe a
variety of interesting unusual phenomena that have been reported to
characterize the decision making properties of real human beings.

The pivotal point of the approach, formalized in QDT, is that
mathematically it is based on the von Neumann theory of quantum
measurements \cite{13}. The formal relation of the von Neumann
measurement theory to quantum information processing has been
considered by \cite{72}. QDT generalizes the quantum measurement
theory to be applicable not merely to simple actions, but also to
composite prospects, which is of paramount importance for the
appearance of decision interference. The principal difference of QDT
from the measurement theory is the existence of a specific {\it
strategic state} characterizing each particular decision maker.

A brief account of the axiomatics of QDT has been published in the
recent letter \cite{12}. The principal scheme of functioning of a
thinking quantum system, imitating the process of decision making,
has been advanced \cite{73}. The applicability of the suggested
quantum approach for analyzing the phenomena of dynamic
inconsistency has been illustrated \cite{74}. The aim of the present
survey is to provide a detailed explanation of the theory and to
demonstrate that it can be successfully applied to real-life
problems of decision making. We also show that the method of
conditional entropy maximization, which is equivalent to the
minimization of an information functional, yields an explicit
relation between the quantum decision theory and the classical
decision theory based on the standard notion of expected
utility.

\section{Mathematical Foundation of Quantum Decision Theory}

In order to formulate in precise mathematical terms the scheme of
information processing and decision making in quantum decision
theory, it is necessary to introduce several definitions. To better
understand these definitions, we shall give some very simple
examples, although much more complicated cases can be invented. The
entity concerned with the decision making task can be a single
human, a group of humans, a society, a computer, or any other system
that is able or enables to make decisions. Throughout the paper, for
the operations with intended actions, we shall use the notations
that are accepted in the literature on decision theory
\cite{58,59,60,61,62,63,64,65,66,67,68,69} and for the physical
states, we shall employ the Dirac notations widely used in quantum
theory \cite{75}.

\vskip 5mm
{\it Definition 1. Action ring}

The process of taking decisions implies that one is deliberating
between several admissible actions with different outcomes,
in order to decide which of the intended actions to choose.
Therefore, the first element arising in decision theory is an
{\it intended action} $A$.

An intended action which, for brevity, can be called an intention or
an action, is a particular thought about doing something. Examples
of intentions could be as follows: ``I would like to marry'' or ``I
would like to be rich'' or ``I would like to establish a firm''.
There can be a variety of intentions, which we assume to be
enumerated by an index $i=1,2,3,\ldots, N$, where the total number
$N$ of actions can be finite or infinite. 

The whole family of all these actions forms the {\it action set} \be
\label{1} \cA \equiv \{ A_i : \; i=1,2,\ldots , N \} \; . \ee The
elements of this set are assumed to be endowed with two binary
operations, addition and multiplication, so that, if $A$ and $B$
pertain to ${\cal A}$, then $AB$ and $A+B$ also pertain to ${\cal
A}$. The addition is associative, such that $A+(B+C)=(A+B)+C$, and
reversible, in the sense that $A+B=C$ implies $A = C-B$. The
multiplication is distributive, $A(B+C)=AB+AC$. The multiplication
is not necessarily commutative, so that, generally, $AB$ is not the
same as $BA$.

Among the elements of the action set (\ref{1}), there is an identity
action $1$, for which $A1 = 1A = A$. The identity action $1$ is not
to ``do nothing'', since inaction is actually an action. This is
well recognized for instance in the field of risk management.
Consider for instance the famous quotes: ``The man who achieves
makes many mistakes, but he never makes the biggest mistake of
all---doing nothing'' (Benjamin Franklin), or ``Life is inherently
risky. There is only one big risk you should avoid at all costs, and
that is the risk of doing nothing'' (Denis Waitley). This also
resonates with the standard recommendations in risk management: ``If
you do not actively attack risks, they will attack you'' or ``Risk
prevention is cheaper than reconstruction.'' Thus, ``not acting'' is
not the identity action.  We interpret the identity action $1$ as
the action of keeping running the present action an individual is
involved in. For instance, if action $A$ is ``to  marry someone'',
the action $1A$ is to marry someone and to confirm this action. The
action $A1$ can be interpreted as first ``being open to decide an
action'' and then to ``decide to marry someone''.

And there exists an impossible action $0$, for which $A0 = 0A = 0$.
Two actions are called disjoint, when their joint action is impossible,
giving $AB = BA = 0$. The action set (\ref{1}), with the described structure,
is termed the {\it action ring}.

We recall that, in mathematics, a ring is an algebraic structure
consisting of a set together with two binary operations (usually called addition
and multiplication), where each operation combines two elements to form a third
element (closure property). This closure property here embodies the fact that
choosing between alternative actions or combining several actions still correspond
to actions.

In the algebra of the elements of the action ring, the meaning of
the operations of addition and multiplication is the same as is
routinely used in the literature
\cite{58,59,60,61,62,63,64,65,66,67,68,69}. The sum $A + B$ means
that either the action $A$ or action $B$ is intended to be realized.
And the product of actions $AB$ implies that both these actions are
to be accomplished together. Instead of writing the sum $A_1 + A_2 +
\cdots$, it is often convenient to use the shorter summation symbol
$\bigcup_i A_i \equiv A_1 + A_2 + \cdots$, which is also the
standard abbreviation. Similarly, for a long product $A_1 A_2
\cdots$, it is convenient to use the shorter notation $\bigcap_i A_i
\equiv A_1 A_2 \cdots$. The use of these standard notations can lead
to no confusion.

\vskip 5mm
{\it Definition 2. Action modes}

\vskip 2mm An action is simple, when it cannot be decomposed into
the sum of other actions. An action is composite, when it can be
represented as a sum of several other actions. If an action is
represented as a sum \be \label{2} A_i =
\bigcup_{\mu=1}^{M_i} A_{i\mu}   \; , \ee whose terms are mutually
incompatible, then these terms are named the {\it action modes}.
Here, $M_i$ denotes the number of modes in action $A_i$. The modes
correspond to different possible ways of realizing an action.
According to the meaning emphasized above, the summation symbol in
Equation (\ref{2}) implies that one of the actions is intended to be
realized.

Action representations, or action modes, are concrete
implementations of an intended action. For instance, the intention
``to marry'' can have as representations the following variants:
``to marry $A$'' or ``to marry $B$", and so on. The intention ``to
be rich'' can have as representations ``to be rich by working hard''
or ``to be rich by becoming a bandit''. The intention ``to establish
a firm" can have as representations ``to establish a firm producing
cars'' or ``to establish a firm publishing books'' and so on. We
number all representations of an $i$-intended action by the index
$\mu=1,2,3,\ldots$. Note that intention representations may include
not only positive intention variants ``to do something'' but also
negative variants such as ``not to do something''. For example,
Hamlet's hesitation ``to be or not to be'' is the intended action
consisting of two representations, one positive and the other
negative.

\vskip 5mm
{\it Definition 3. Elementary prospects}

\vskip 2mm
Generally, decision taking is not necessarily associated with
a choice of just one action among several simple given actions,
but it involves a choice between several complex actions. The
simplest such complex action is defined as follows. Let the
multi-index $n = \{\nu_1, \nu_2,...,\nu_N\}$ be a set of
indices enumerating several chosen modes, under the condition
that each action is represented only by one of its modes. The
{\it elementary prospect} is the conjunction
\be
\label{3}
e_n \equiv \bigcap_{i=1}^N A_{i\nu_i}   \; ,
\ee
of the chosen modes, one for each of the actions from the action
ring (\ref{1}). The total set of all elementary prospects will
be denoted as $\{e_n\}$.

\vskip 5mm
{\it Definition 4. Composite prospects}

\vskip 2mm
A prospect is composite, when it cannot be represented as an
elementary prospect (\ref{3}). Generally, a composite prospect
is a conjunction
\be
\label{4}
\pi_j = \bigcap_n A_{j_n}
\ee
of several composite actions of form (\ref{2}), where each of
the factors $A_{j_n}$ pertains to the action ring (\ref{1}).
While expression (\ref{4}) has the similar form as (\ref{3}),
the difference is that the actions $A_{j_n}$ in (\ref{4}) are
composite while the actions $A_{i\nu_i}$ in (\ref{3}) are elementary
action modes.

A prospect is a set of several intended actions or several intention
representations. In reality, a decision maker is always motivated by a
variety of intentions, which are mutually interconnected. Even the
realization of a single intention always involves taking into account
many other related intended actions.

\vskip 5mm
{\it Definition 5. Prospect lattice}

\vskip 2mm
All possible prospects, among which one needs to make a choice, form a set
\be
\label{5}
\cL = \{ \pi_j : \; j=1,2,\ldots,N_L \} \; .
\ee
The set is assumed to be equipped with the binary relations
$ >, <, =, \geq, \leq$, so that each two prospects $\pi_i$ and
$\pi_j$ in $\cal L$ are related as either  $\pi_i>\pi_j$, or
$\pi_i=\pi_j$, or $\pi_i\geq\pi_j$, or $\pi_i< \pi_j$, or
$\pi_i\leq \pi_j$. For a while, it is sufficient to assume that
such an ordering exists. Then, the ordered set (\ref{5}) is called
a {\it lattice}. The explicit ordering procedure associated with
decision making will be given below.

\vskip 5mm
{\it Definition 6. Mode space}

\vskip 2mm
To each action mode $A_{i\mu}$, there corresponds the {\it mode
state} $|A_{i\mu}\rgl$, which is a complex function $\cA\ra\cC$,
and its Hermitian conjugate $\lgl A_{i\mu}|$. Here we employ the
Dirac notation \cite{75}. We assume that a scalar product is defined,
such that the mode states, pertaining to the same action, are
orthonormalized:
\be
\label{6}
\lgl A_{i\mu} | A_{i\nu} \rgl = \dlt_{\mu\nu} \; .
\ee
The {\it mode space} is the closed linear envelope
\be
\label{7}
\cM_i \equiv {\rm Span} \{ | A_{i\mu}\rgl : \;
\mu = 1,2, \ldots, M_i\} \; ,
\ee
spanning all mode states. By this definition, the mode space,
corresponding to an $i$-action $A_i$, is a Hilbert space of
dimensionality $M_i$. The elements of the mode space will be called
the {\it intention states}.

\vskip 5mm
{\it Definition 7. Mind space}

\vskip 3mm
To each elementary prospect $e_n$, there corresponds the {\it basic
state} $|e_n\rgl$, which is a complex function $\cA^N\ra\cC$, and
its Hermitian conjugate $\lgl e_n|$. The structure of a basic state
is
\be
\label{8}
| e_n\rgl \equiv | A_{1\nu_1} A_{2\nu_2} \ldots
A_{N\nu_N} \rgl = \bigotimes_{i=1}^N | A_{i\nu_i}\rgl  \; .
\ee
The scalar product is assumed to be defined, such that the basic
states are orthonormalized:
\be
\label{9}
\lgl e_m | e_n \rgl = \prod_{i=1}^N \dlt_{\mu_i\nu_i}
\equiv \dlt_{mn} \; .
\ee
The {\it mind space} is the closed linear envelope
\be
\label{10}
\cM \equiv {\rm Span} \{ | e_n\rgl \} =
\bigotimes_{i=1}^N \cM_i \;  ,
\ee
spanning all basic states (\ref{8}). Hence, the mind space is a
Hilbert space of dimensionality
$$
{\rm dim} \cM = \prod_{i=1}^N M_i \; .
$$
The vectors of the mind space represent all possible actions and
prospects considered by a decision maker.

The family of the basic states forms the {\it mind basis} $\{ |e_n>\}$
in the mind space. Different states belonging to the mind basis are
assumed to be disjoint, in the sense of being orthogonal. Since the
modulus of each state has no special meaning, these states are also
normalized to one. This is formalized as the orthonormality of the basis.

{\it Definition 8. Prospect states}

\vskip 2mm To each prospect $\pi_j$, there corresponds a state
$|\pi_j\rgl\in\cM$ that is a member of the mind space (\ref{10}).
Hence, the prospect state can be represented as an expansion over
the basic states \be \label{11} |\pi_j \rgl = \sum_n \; a_{jn} | e_n
\rgl \; . \ee The expansion coefficients in Equation (\ref{11}) are
assumed to be defined by the decision maker, so that $|a_{jn}|^2$
gives the weight of the state $|e_n>$ into the general prospect.

The prospects are enumerated with the index $j=1,2,\ldots$. The total
set $\{|\pi_j>\}$ of all prospect states $|\pi_j>$, corresponding to
all admissible prospects, forms a subset of the space of mind. The set
$\{|\pi_j>\}\subset\cM$ can be called the prospect-state set.

The prospect states are not required to be mutually orthogonal and
normalized to one, so that the scalar product
$$
\lgl \pi_i | \pi_j \rgl = \sum_n \; a_{in}^* a_{jn}
$$
is not necessarily a Kronecker delta. The normalization condition will
be formulated for the prospect probabilities to be defined below.

The fact that different prospect states are not necessarily
orthogonal assumes that the related prospects are not necessarily
incompatible. The incompatibility is supposed only for the
elementary prospects (\ref{3}), whose states form the
basis in the mind space (\ref{10}) and are orthogonal to each other,
according to Equation (\ref{9}). But an arbitrarily defined
composite prospect, generally, is not required to be orthogonal to
all other considered prospects. In particular, this can be so, but,
in general, we do not need this property.

The prospect states are not normalized to one, since, imposing such a
condition would over define these states. The normalization condition
will be imposed below on the prospect probabilities. Generally,
imposing two normalization conditions could make them inconsistent with
each other. So, we need just one normalization condition for the prospect
probabilities, which is necessary for the correct definition of
the related probability measure.

Being, generally, not orthonormalized, the prospect states do not form a
basis in the mind space.

\vskip 5mm
{\it Definition 9. Strategic state}

\vskip 2mm
Among the states of the mind space, there exists a special fixed
state $|s\rgl\in\cM$, playing the role of a reference state, which
is termed the {\it strategic state}. The strategic state of mind is a
fixed vector characterizing a particular decision maker, with his/her
beliefs, habits, principals, etc., that is, describing each decision
maker as a unique subject. Hence, each space of mind possesses a
unique strategic state. Different decision makers possess different
strategic states.

Being in the mind space (\ref{10}), this state can be represented as
the decomposition \be \label{12} | s \rgl = \sum_n c_n | e_n \rgl \;
. \ee Being a unique state, characterizing each decision maker like
its fingerprints, it can be normalized to one: \be
\label{13} \lgl s | s \rgl = 1 \; . \ee From Equations (\ref{12})
and (\ref{13}), it follows that
$$
\sum_n | c_n|^2 = 1   .
$$
The existence of the strategic state, uniquely defining each
particular decision maker, is the principal point distinguishing
an {\it active thinking} quantum system from a {\it passive} quantum
system subject to measurements from an external observer. For a passive
quantum system, predictions of the outcome of measurements are
performed by summing (averaging) over all possible statistically
equivalent states, which can be referred to as a kind of
``annealed''  situation. In contrast, decisions and observations
associated with a thinking quantum system occur in the presence
of this unique strategic state, which can be thought of as a kind
of fixed ``quenched'' state. As a consequence, the outcomes of the
applications of the quantum-mechanical formalism will thus be
different for thinking versus passive quantum systems.

\vskip 5mm
{\it Definition 10. Prospect operators}

\vskip 2mm
Each prospect state $|\pi_j\rgl$, together with its Hermitian
conjugate $\lgl\pi_j|$, defines the {\it prospect operator}
\be
\label{14}
\hat P(\pi_j) \equiv | \pi_j \rgl \lgl \pi_j | \; .
\ee
By this definition, the prospect operator is self-adjoint. The
family of all prospect operators forms the involutive bijective
algebra that is analogous to the algebra of local observables in
quantum theory. Since the prospect states, in general, are neither
mutually orthogonal nor normalized, the squared operator
$$
\hat P^2(\pi_j) = \lgl \pi_j | \pi_j \rgl \hat P(\pi_j)
$$
contains the scalar product
$$
 \lgl \pi_j | \pi_j \rgl = \sum_n | a_{nj}|^2 \;  ,
$$
which does not equal to one. This tells us that the prospect
operators, generally, are not idempotent, thus, they are not
projection operators. It is only when the prospect is elementary
that the related prospect operator
$$
\hat P(e_n) = | e_n \rgl \lgl e_n |
$$
becomes idempotent and is a projection operator. But, in general,
this is not so.

The properties of the prospect operators follow immediately from
those of the prospect states and definition (\ref{14}). Recall that
the prospect operators are analogous to the operators of local
observables in quantum theory. The latter operators are not required
to be idempotent. So, the prospect operators are also not required
to be such. The intuition of why the prospect operators are not
idempotent could be justified by understanding that, in general,
a prospect realized twice results in the consequences that are not
necessarily the same as a sole prospect realization. For instance,
to marry twice is not the same as to marry once.

\vskip 5mm
{\it Definition 11. Prospect probabilities}

\vskip 2mm In quantum theory, the averages over the system state,
for the operators from the algebra of local observables, define the
observable quantities. In the same way, the averages, over the
strategic state, for the prospect operators define the observable
quantities, the {\it prospect probabilities} \be \label{15} p(\pi_j)
\equiv \lgl s | \hat P(\pi_j) | s \rgl  \; . \ee These are assumed
to be normalized to one: \be \label{16} \sum_{j=1}^{N_L}\; p(\pi_j)
= 1 \; , \ee where the summation is over all prospects from the
prospect lattice (\ref{5}). By their definition, the 
quantities (\ref{15}) are non-negative, since Equation (\ref{15})
reduces to the modulus of the transition amplitude
squared
$$
p(\pi_j) = | \lgl \pi_j | s \rgl |^2  \; .
$$
The normalization in Equation (\ref{16}) is necessary for the set
$\{p(\pi_j)\}$ be the scalar probability measure. In plane words,
the fact that all prospects probabilities are summed to one implies
that one of them is to be certainly realized.

\vskip 5mm
{\it Definition 12. Utility factor}

\vskip 2mm
The diagonal form
\be
\label{17}
p_0(\pi_j) \equiv \sum_n \;
\lgl s | \hat P(e_n) \hat P(\pi_j) \hat P(e_n) | s \rgl
\ee
plays the role of the expected utility in classical decision making,
justifying its name as the {\it utility factor}. In order to
be generally defined and to be independent of the chosen units of
measurement, the utility factor (\ref{17}) can be normalized as
\be
\label{18}
\sum_{j=1}^{N_L} \; p_0(\pi_j) = 1 \; .
\ee
The fact that the utility factor (\ref{17}) is really equivalent to the
classical expected utility follows from noticing that
$$
\hat P(e_n) | s \rgl = c_n | e_n \rgl \; ,
$$
hence Equation (\ref{17}) acquires the form
$$
p_0(\pi_j) = \sum_n | c_n|^2 \lgl e_n | \hat P(\pi_j) | e_n \rgl \; ,
$$
where $<e_n|\hat{P}(\pi_j)|e_n>$ plays the role of a utility function,
weighted with the probability $|c_n|^2$.

\vskip 5mm
{\it Definition 13. Attraction factor}

\vskip 2mm The nondiagonal term \be \label{19} q(\pi_j) \equiv
\sum_{m\neq n} \; \lgl s | \hat P(e_m) \hat P(\pi_j) \hat P(e_n) | s
\rgl \ee corresponds to the quantum interference effect. Its
appearance is typical of quantum mechanics. Such nondiagonal terms
do not occur in classical decision theory. This term can be called
the {\it interference factor}. Interpreting its meaning in decision
making, we can associate its appearance as resulting from
the system deliberation between several alternatives, when
deciding which of the latter is more attractive. Thence, the name
``attraction factor''. Using expansion (\ref{12}) in Equation
(\ref{19}) yields
$$
q(\pi_j) \equiv \sum_{m\neq n} c_m^* c_n \lgl e_m |
\hat P(\pi_j) | e_n \rgl \; ,
$$
which shows that the interference occurs between different
elementary prospects in the process of considering a composite
prospect $\pi_j$. It is worth stressing that the interference
factor is nonzero only when the prospect $\pi_j$ is composite.
If it were elementary, say $\pi_j=e_k$ then, since
$$
\hat P(e_k) | e_n \rgl = \dlt_{nk} | e_n \rgl \; ,
$$
we would have
$$
q(e_k) = \sum_{m\neq n} c_m^* c_n \dlt_{mn}
\dlt_{nk} = 0 \; ,
$$
and no interference would arise.

Between two prospects, the one which enjoys the larger attraction
factor is more attractive.

\vskip 5mm
{\it Definition 14. Prospect ordering}

\vskip 2mm
In defining the prospect lattice (\ref{5}), we have assumed
that the prospects could be ordered. Now, after introducing
the scalar probability measure, we are in a position to give
an explicit prescription for the prospect ordering. We say that
the prospect $\pi_1$ is {\it preferable} to $\pi_2$ if and only if
\be
\label{20}
p(\pi_1) > p(\pi_2) \qquad ( \pi_1 > \pi_2) \;  .
\ee
Two prospects are called indifferent if and only if
\be
\label{21}
p(\pi_1) = p(\pi_2) \qquad ( \pi_1 = \pi_2) \; .
\ee
And the prospect $\pi_1$ is preferable or indifferent to $\pi_2$
if and only if
\be
\label{22}
p(\pi_1) \geq p(\pi_2) \qquad ( \pi_1 \geq \pi_2) \; .
\ee
These binary relations provide us with an explicit prospect ordering
making the prospect set (\ref{5}) a lattice.

\vskip 5mm
{\it Definition 15. Optimal prospect}

\vskip 2mm
Since all prospects in the lattice are ordered, it  is straightforward
to find among them that one enjoying the largest probability. This
defines the {\it optimal prospect} $\pi_*$ for which
\be
\label{23}
p(\pi_*) \equiv \sup_j p(\pi_j) \; .
\ee
Finding the optimal prospect is the final goal of the decision-making
process. Since the prospect probabilities are non-negative, it is
possible to find the minimal prospect  in the lattice (\ref{5}) with
the smallest probability. And the largest probability defines the
optimal prospect $\pi_*$. Therefore the prospect set (\ref{5}) is
a complete lattice.

\vskip 2mm

{\bf Remark}. Generally speaking, all states of the mind space can depend
on time $t$. We do not write explicitly the time dependence, when this
makes no difference for the considerations developed below. When this is
important, we shall denote the time dependence explicitly.

\section{Entangled Prospect States}

Prospect states can be of two qualitatively different types.
\begin{itemize}
\item
A {\it disentangled prospect state} is a prospect state
which is represented as the tensor product of the intention states:
\be
\label{24}
|f > = \otimes_i |\psi_i > \; .
\ee

\item
An {\it entangled prospect state} is any prospect state that cannot be
reduced to the tensor product form of the disentangled prospect
states (\ref{24}).
\end{itemize}

\vskip 2mm

We define the {\it disentangled set} as the collection of all
admissible disentangled prospect states of form
(\ref{24}): \be \label{25} \cD \equiv \{ |f > = \otimes_i |\psi_i >,
\; |\psi_i
> \in \cM_i\}~. \ee

In quantum theory, it is possible to construct various entangled and
disentangled states (see, e.g., \cite{76,77}). In order
to explain how entangled states appear in the quantum theory of
decision making, let us illustrate the above definitions by an
example of a prospect consisting of two intended actions with two
mode representations each. Let us consider the prospect of the
following two intentions: ``to get married'' and ``to become rich''.
And let us assume that the intention ``to get married" consists of
two representations, ``to marry $A$'', with the mode state $|A>$,
and ``to marry $B$'', with the mode state $|B>$. And let the
intention ``to become rich'' be formed by two representations, ``to
become rich by working hard'', with the mode state $|W>$, and ``to
become rich by being a gangster'', with the mode state $|G>$. Thus,
there are two intention states of the following type: \be \label{26}
|\psi_1 > = a_1|A> + a_2|B>  \; , \qquad |\psi_2 > = b_1|W> + b_2|G>
\; . \ee The general prospect state has the form \be \label{27} |\pi
> = c_{11}|AW> + c_{12}|AG> + c_{21}|BW> + c_{22}|BG> \; , \ee where
the coefficients $c_{ij}$ belong to the field of complex numbers.

Depending on the values of the coefficients $c_{ij}$, the prospect
state (\ref{27}) can be either disentangled or entangled. If it is
disentangled, it must be of the tensor product type (\ref{24}),
which for the present case reads \be \label{28} |f > =
|\psi_1
> \otimes |\psi_2 > = a_1 b_1|AW> + a_1 b_2|AG> + a_2b_1|BW> +
a_2b_2|BG> \; . \ee Both states (\ref{27}) and (\ref{28}) include
four basic states:
\begin{itemize}
\item
``to marry $A$ and to work hard'', $|AW>$,
\item
``to marry $A$ and become a gangster'', $|AG>$,
\item
``to marry  $B$ and to work hard'', $|BW>$,
\item
``to marry $B$ and become a gangster'',  $|BG>$.
\end{itemize}

However, the structure of states (\ref{27}) and (\ref{28}) is
different. The prospect state (\ref{27}) is more general and can be
reduced to state (\ref{28}) for special values of the coefficients
$c_{ij}$, but the opposite may not be possible. For instance, the
prospect state \be \label{29} c_{12}|AG> + c_{21}|BW> \; , \ee which
is a particular example of state (\ref{27}) cannot be reduced to any
of the states (\ref{28}), provided that both coefficients $c_{12}$
and $c_{21}$ are non-zero. In quantum mechanics, this state would be
called the Einstein-Podolsky-Rosen state, one of the most famous
examples of an entangled state \cite{78}. Another example is the
prospect state \be \label{30} c_{11}|AW > + c_{22}| BG > \; , \ee
whose quantum-mechanical analog would be called the Bell state
\cite{79}. In the case where both $c_{11}$ and $c_{22}$ are
non-zero, the Bell state cannot be reduced to any of the states
(\ref{28}) and is thus entangled.

In contrast with the above two examples, the prospect states
$$
c_{11}|AW> + c_{12}|AG> \; ,  \qquad c_{11}|AW> + c_{21}|BW> \; ,
$$
$$
c_{12}|AG> + c_{22}|BG> \; ,  \qquad c_{21}|BW> + c_{22}|BG> \; ,
$$
are {\it disentangled}, since all of them can be reduced to the
form (\ref{28}).

Other examples of {\it entangled} prospect states are
$$
c_{11}|AW> + c_{12}|AG> + c_{21}|BW> \; , \qquad
c_{11}|AW> + c_{12}|AG> + c_{22}|BG> \; ,
$$
$$
c_{11}|AW> + c_{21}|BW> + c_{22}|BG> \; , \qquad
c_{12}|AG> + c_{21}|BW> + c_{22}|BG> \; ,
$$
where all coefficients are assumed to be non-zero.

Since the coefficients $c_{ij}=c_{ij}(t)$ are, in general, functions
of time, it may happen that a prospect state at a particular time is
entangled, but becomes disentangled at another time or, vice versa,
a disentangled prospect state can be transformed into an entangled
state with changing time \cite{80}.

The state of a human being is governed by its physiological
characteristics and the available information \cite{81,82}. These
properties are continuously changing in time. Hence the strategic
state (\ref{12}), specific of a person at a given time,
may also display temporal evolution, according to different
homeostatic processes adjusting the individual to the changing
environment \cite{83}.

\section{Procedure of Decision Making}

The process of quantum decision making possesses several features that
make it rather different from the classical decision making. These main
features are emphasized below.

\subsection{Probabilistic Nature of Decision Making}

Quantum decision making is described as an intrinsically probabilistic
procedure. The first step consists in evaluating, consciously and/or
subconsciously, the probabilities of choosing different prospects from
the point of view of their usefulness and/or appeal to the choosing agent.
The strategic state of mind of an agent at some time $t$ is represented by
the state $|s>$. Then, the probability of realizing a prospect $\pi_j$
with the prospect state $|\pi_j>$, under the given strategic state $|s>$,
characterizing the agent's state of mind at that time, according to
Definition 11, is the {\it prospect probability} $p(\pi_j)$. The prospect
probabilities, defined in (\ref{15}), possess all the standard probability
properties, with the normalization condition (\ref{16}).

The probabilities are defined in Equation (\ref{15}) through the
prospect states and the strategic state of mind. The latter is
normalized to one, according to Equation (\ref{13}). By their
definition, the probabilities are summed to one, as in Equation
(\ref{16}). But the prospect states do not need to be normalized to
one, as is stressed in Definition 8. This means that different
prospects can have, and usually do have, different weights,
corresponding to their different probabilities. In quantum physics,
this situation would be similar to defining the cross-section in a
scattering experiment over a system containing elementary particles
and composite clusters (prospects) formed by these particles.

In the traditional theory of decision making, based on the utility
function, the optimal decision corresponds, by definition, to the
maximal expected utility which is associated with the maximal
anticipated usefulness and profit resulting from the chosen action.
In contrast, QDT recognizes that the behavior of an individual is
probabilistic, not deterministic. The prospect probability
(\ref{15}) quantifies the probability that a given individual
chooses the prospect $\pi_j$, under his/her strategic state of mind
$|s>$ at a given time $t$. This translates in experiments into a
prediction on the frequency of the decisions taken by an ensemble of
subjects under the same conditions. The observed frequencies of
different decisions taken by an ensemble of non-interacting subjects
making a decision under the same conditions serve as the observable
measure of the subjective probability. It is, actually, well-known
that subjective probabilities can be calibrated by frequencies or
fractions \cite{84,85}.

This specification also implies that the same subject, prepared
under the same conditions with the same entangled strategic state of
mind $|s>$ at two different times, may choose two different
prospects among the same set of prospects, with different relative
frequencies determined by the corresponding prospect probabilities
(\ref{15}). Verifying this prediction is a delicate empirical
question, because of the possible impact of the ``memory'' of the
past decisions on the next one. In order for the prediction to hold,
the two repetitions of the decision process should be independent.
Otherwise, the strategic state of mind in the second experiment
keeps a memory of the previous choice, which biases the results.
This should not be confused with the fact that the projection of the
strategic state of mind onto the prospect state $\pi_j$, when the
decision is made to realize this prospect, ensures that the
individual will in general keep his/her decision, whatever it is,
when probed a second time sufficiently shortly after the first
decision so that the strategic state of mind, realized just after
the projection, has not had time yet to evolve appreciably.

In QDT, the concept of an optimal decision is replaced by a
probabilistic decision, when the prospect, which makes $p(\pi_j)$
given by (\ref{15}) maximal, is the one which corresponds best to
the given strategic state of mind of the decision maker. In that
sense, the prospect that makes $p(\pi_j)$ maximal can be called
optimal with respect to the strategic state of mind. Using the
mapping between the subjective probabilities and the frequentist
probabilities observed on ensemble of individuals, the prospect that
makes $p(\pi_j)$ maximal, will be chosen by more individuals that
any other prospect, in the limit of large population sampling sizes.
However, other less probable prospects will also be
chosen by some smaller subset of the population with frequencies
given by the corresponding quantum mechanical probabilities given
above.

\subsection{Entangled Decision Making}

As is explained above, a prospect state $|\pi_j >$ does not have in
general the form of the product (\ref{24}), which means that it is
entangled. Therefore, the prospect probability $p(\pi_j)$, generally,
cannot be reduced to a product:
$$
p(\pi_j) \neq \prod_i p(A_i) \; .
$$
In other words, usually the decision making process is naturally
entangled.

Consider the example of the specific prospect state (\ref{27})
associated with the two intentions ``to get married'' and ``to
become rich''. And suppose that $A$ does not like gangsters, so that
it is impossible to marry $A$ and at the same time being a gangster.
This implies that the prospect $AG$ cannot be realized, hence
$c_{12}=0$. Assume that $B$ dreams of becoming rich as fast as
possible, and a gangster spouse is much more luring for $B$ than a
dull person working hard, which implies that $c_{21}=0$. In this
situation, the prospect state (\ref{27}) reduces to the entangled
Bell state $c_{11}|AW>+c_{22}|BG>$. A decision performed under these
conditions, resulting in an entangled state, is entangled.

\subsection{Non-commutativity of Decisions and History Dependence}

There exist numerous real-life examples when decision makers fail to
follow their plans and change their mind simply because they
experience different outcomes on which their intended plans were
based. This change of plans after experiencing particular outcomes
is the effect known as dynamic inconsistency
\cite{86,87,88}. In our language \cite{74}, this can be considered
as a consequence of the non-commutativity of subsequent decisions,
resulting from interference and entanglement between intention
representations. After studying the effect of interference, we shall
give in what follows a rigorous mathematical formulation of the
non-commutativity of decisions.

\section{Interference of Intended Actions}

Interference is the effect that is typical of all those phenomena
which are described by wave equations. Following the Bohr's idea
\cite{36,37,38,39} of describing mental processes in terms of
quantum mechanics, one is immediately confronted with the
interference effect, since the physical states in quantum mechanics
are characterized by wave functions. The possible occurrence of
interference in the problems of decision making has been discussed
before on different grounds (see, e.g., \cite{89}). However, no
general theory has been suggested, which would explain why and when
such a kind of effect would appear, how to predict it, and how to
give a quantitative analysis of it that can be compared with
empirical observations. In our approach, interference in decision
making arises only when one takes a decision involving composite
intentions. The corresponding mathematical treatment of these
interferences within QDT is presented in the following subsections.

\subsection{Illustration of Interference in Decision Making}

As an illustration, let us consider the following situation of two
intended actions, ``to get a friend'' and ``to become rich''. Let
the former intention have two mode representations ``to get a friend
$A$" and ``to get a friend $B$''. And let the second intention also
have two representations, ``to become rich by working hard" and ``to
become rich by being a gangster''. The corresponding strategic mind
state is given by Equation (\ref{12}), with the evident notation for
the basic states $|e_n>$ and the coefficients $c_n$ represented by
the identities \be \label{31} c_{11} \equiv c_{AW} \; , \qquad
c_{12} \equiv c_{AG} \; , \qquad c_{21} \equiv c_{BW} \; , \qquad
c_{22} \equiv c_{BG} \; . \ee

Suppose that one does not wish to choose between these two friends
in an exclusive manner, but one hesitates of being a friend to $A$
as well as $B$, with the appropriate weights. This means that one
considers the intended actions $A$ and $B$, while the way of life,
either to work hard or to become a gangster, has not yet been decided.

The corresponding composite prospects
\be
\label{32}
\pi_A = A (W + G) \; , \qquad \pi_B = B (W + G)
\ee
are characterized by the prospect states
\be
\label{33}
|\pi_A > \; = \; a_1|AW > + a_2|AG > \; , \qquad
|\pi_B > \; = \; b_1|BW > + b_2|BG > \; .
\ee
Let us stress that the weights correspond to the intended actions,
among which the choice is yet to be made. And one should not confuse the
intended actions with the actions that have already been realized. One
can perfectly deliberate between keeping this or that friend, in the same
way, as one would think about marrying $A$ or $B$ in another example above.
This means that the choice has not yet been made. And before it is made,
there exist deliberations involving stronger or weaker intentions to both
possibilities. Of course, one cannot marry both (at least in most Christian
communities). But before marriage, there can exist the dilemma between
choosing this or that individual.

Calculating the scalar products
\be
\label{34}
<\pi_A|s > \; = \; a_1^* c_{11} + a_2^* c_{12} \; ,
\qquad
<\pi_B|s > \; = \; b_1^* c_{21} + b_2^* c_{22} \; ,
\ee
we find the prospect probabilities.
\be
\label{35}
p(\pi_A) = \left | a_1^* c_{11} + a_2^* c_{12} \right |^2 \; ,
\qquad
p(\pi_B) = \left | b_1^* c_{21} + b_2^* c_{22} \right |^2 \; .
\ee

Recall that the prospects are characterized by vectors pertaining to
the space of mind $\cM$, which are not necessarily normalized to one or
orthogonal to each other. The main constraint is that the total set of
prospect states $\{|\pi_j>\}$ be such that the related probabilities
$$
p(\pi_j)\equiv|<\pi_j|s>|^2
$$
be normalized to one, according to the normalization condition
(\ref{16}).

The probabilities (\ref{35}) can be rewritten in another form by
introducing the partial probabilities
$$
p(AW) \equiv | a_1 c_{11}|^2 \; , \qquad
p(AG) \equiv | a_2 c_{12}|^2 \; ,
$$
\be
\label{36}
p(BW) \equiv | b_1 c_{21}|^2 \; , \qquad
p(BG) \equiv | b_2 c_{22}|^2 \; ,
\ee
and the interference terms
\be
\label{37}
q(\pi_A) \equiv 2{\rm Re} \left ( a_1^* c_{11} a_2 c_{12}^*
\right ) \; , \qquad
q(\pi_B) \equiv 2{\rm Re} \left ( b_1^* c_{21} b_2 c_{22}^*
\right ) \; .
\ee
Then the probabilities (\ref{35}) become
\be
\label{38}
p(\pi_A) = p(AW) + p(AG) + q(\pi_A) \; , \qquad
p(\pi_B) = p(BW) + p(BG) + q(\pi_B)  \; .
\ee

Let us define the {\it uncertainty angles}
\be
\label{39}
\Dlt(\pi_A) \equiv {\rm arg} \left ( a_1^* c_{11} a_2 c_{12}^*
\right ) \; ,\qquad
\Dlt(\pi_B) \equiv {\rm arg} \left ( b_1^* c_{21} b_2 c_{22}^*
\right )
\ee
and the {\it uncertainty factors}
\be
\label{40}
\vp(\pi_A) \equiv \cos \Dlt(\pi_A) \; , \qquad
\vp(\pi_B) \equiv \cos \Dlt(\pi_B)\;  .
\ee
Using these, the interference terms (\ref{37}) take the form
\be
\label{41}
q(\pi_A) = 2\vp(\pi_A) \; \sqrt{p(AW) p(AG) } \; , \qquad
q(\pi_B) = 2\vp(\pi_B) \; \sqrt{p(BW) p(BG) } \;  .
\ee
The interference terms characterize the existence of deliberations between
the decisions of choosing a friend and, at the same time, a type of work.

This example illustrates the observation that the phenomenon of decision
interference appears when one considers a composite prospect with several
intention representations assumed to be realized simultaneously.
Treating a composite prospect as a combination of several sub-prospects,
we could consider the global decision as a collection of sub-decisions.
Then the arising interference would occur between these sub-decisions. From
the mathematical point of view, it appears more convenient to combine several
sub-decisions into one global decision and to analyze the interference of
different intentions. Thus, we can state that interference in decision
making appears only when one decides about a composite prospect.

For the above example of decision making in the case of two
intentions, ``to get a friend'' and ``to be rich'', the appearance
of the interference can be understood as follows. In real life, it
is too problematic, and practically impossible, to become a very
close friend to several persons simultaneously, since conflict of
interests often arises between the friends. For instance, doing a
friendly action to one friend may upset or even harm another friend.
Any decision making, involving mutual correlations between two
persons, necessarily requires taking into account their, sometimes
conflicting, interests. This is, actually, one of the origins of the
interference in decision making. Another powerful origin of
intention interference is the existence of emotions, as will be
discussed in the following sections.

\subsection{Conditions for Interference Appearance}

The situations for which intention interferences is impossible can be
classified into two cases, which are examined below. From this
classification, we conclude that the necessary conditions for the
appearance of intention interferences are that the dimensionality of mind
should be not lower than two and that there should be some uncertainty in
the considered prospect.

\vskip 3mm
{\bf One-dimensional mind}

\vskip 2mm
Suppose there are several intended actions $\{ A_i\}$, enumerated by the
index $i=1,2,\ldots$, whose number can be arbitrary. But each intention
possesses only a single representation $|A_i>$. Hence, the dimension
of ``mind''  as defined in Definition 7, is $dim\cM=1$. Only a single
basic vector exists, which forms the strategic state
\be
\label{42}
 |s> = | A_1 A_2 \ldots > \; = \; \otimes_i \; |A_i> \; .
\ee
In this one-dimensional mind, all prospect states are disentangled, being
of the type
\be
\label{43}
|\pi > = c\; | A_1 A_2 \ldots > \qquad (|c|=1) \; .
\ee
Therefore, only one probability exists:
\be
\label{44}
p = | < \pi|s > |^2 = 1 \; .
\ee

Thus, despite the possible large number of arbitrary intentions,
they do not interfere, since each of them has just one representation.
There can be no intention interference in one-dimensional mind.

\vskip 3mm
{\bf Absence of uncertainty}

\vskip 2mm Another important condition for the appearance of
intention interference is the existence of uncertainty. To
understand this statement, let us consider a given mind with a large
dimensionality $dim\cM >1$, characterized by a strategic state
$|s>$. Let us analyze a certain prospect with the state \be
\label{45} |\pi > = c |s> \qquad (|c|=1) \; , \ee with an arbitrary
strategic state $|s>$. Then again, the corresponding prospect
probability is the same as in Equation (\ref{44}), and no
interference can arise.

Thus, the necessary conditions for the appearance of interference are the
existence of uncertainty and the dimensionality of mind not lower
than $2$.

\subsection{Interference Alternation}

The interference terms, forming the attraction factor (\ref{19}), enjoy a
very important property that can be called the {\it theorem of interference
alternation}.

\vskip 2mm

{\bf Theorem 1}: {\it The process of decision making, associated with the
prospect probabilities (\ref{15}) and occurring under the normalization
conditions (\ref{16}) and (\ref{18}), is characterized by the alternating
interference terms, such that the sum of all attraction factors vanishes}:
\be
\label{46}
\sum_j q(\pi_j) = 0 \; .
\ee

\vskip 2mm

{\it Proof}: The proof follows directly from Definitions 13, 15, and 19,
taking into account the normalization conditions (\ref{16}) and (\ref{18}).

\vskip 2mm

In order to illustrate in more detail the meaning of the above theorem,
let us consider a particular case of two intentions, one composing a
set $\{ A_i\}$ of $M_1$ representation modes, and another one forming
a set $\{ X_j\}$ of $M_2$ modes. The total family of intended actions is
therefore
$$
\{ A_i,X_j|\; i=1,2,\ldots,M_1; \; j=1,2,\ldots,M_2\} \; .
$$
The basis in the mind space is the set $\{|A_iX_j>\}$. The strategic state
of mind can be written as an expansion over this basis,
\be
\label{47}
|s> = \sum_{ij} c_{ij} |A_iX_j> \; ,
\ee
with the coefficients satisfying the standard normalization
$$
\sum_{ij} |c_{ij}|^2 = 1 \; .
$$

Let us assume that we are mainly interested in the representation set
$\{ A_i\}$, while the representations from the set $\{ X_j\}$ are treated
as secondary. A prospect that is formed of a fixed intention
representation $A_i$, and which can be realized under the occurrence of
any of the representations $X_j$, corresponds to the prospect state
\be
\label{48}
|A_i X > = \sum_j \al_{ij}|A_i X_j> \; ,
\ee
where $X = \bigcup_j X_j$. The probability of realizing the considered
prospect is
\be
\label{49}
p(A_i X) \equiv | <A_i X|s >|^2 \; ,
\ee
according to Definition 11.

Following the above formalism of describing the intention
interferences, we use the notation \be \label{50} p(A_i X_j) \equiv
|\al_{ij} c_{ij}|^2 \ee for the joint probability of $A_i$ and
$X_j$; and we denote the partial interference terms as \be
\label{51} q_{jk}(A_i X) \equiv 2{\rm Re}\left ( \al_{ij}^* c_{ij}
c_{ik}^* \al_{ik} \right ) \; . \ee Then, the probability of $A_i
X$, given by Equation (\ref{48}), becomes \be \label{52} p(A_i X) =
\sum_j p(A_i X_j) + \sum_{j<k} q_{jk}(A_i X) \; . \ee

The interference terms appear due to the existence of uncertainty.
Therefore, we may define the {\it uncertainty factor}
\be
\label{53}
\vp_{jk}(A_i X) \equiv \cos\Dlt_{jk}(A_i X) \; ,
\ee
where the uncertainty angle is
$$
\Dlt_{jk}(A_i X) \equiv
{\rm arg}(\alpha_{ij}^* c_{ij}c_{ik}^*\alpha_{ik}) \; .
$$
Then, the interference term (\ref{51}) takes the form
\be
\label{54}
q_{jk}(A_i X) = 2\vp_{jk}(A_i X) \;
\sqrt{p(A_iX_j) \; p(A_iX_k)} \; .
\ee
The attraction factor (\ref{19}) here is nothing but the sum of the
interference terms:
\be
\label{55}
q(A_i X) \equiv \sum_{j<k} q_{jk}(A_i X) \; .
\ee
This allows us to rewrite probability (\ref{52}) as
\be
\label{56}
p(A_i X) = \sum_j p(A_i X_j) + q_(A_i X) \; .
\ee

The joint and conditional probabilities are related in the standard
way
\be
\label{57}
p(A_iX_j) = p(A_i|X_j) p(X_j) \; .
\ee

We assume that the family of intended actions is such that at least one
of the representations from the set $\{ A_i\}$ has to be certainly
realized, which means that
\be
\label{58}
\sum_i p(A_i X) = 1 \; ,
\ee
and that at least one of the representations from the set $\{ X_j\}$ also
necessarily happens, that is,
\be
\label{59}
\sum_j p(X_j) = 1 \; .
\ee
Along with these conditions, we keep in mind that at least one of the
representations from the set $\{ A_i\}$ must be realized for each given
$X_j$, which implies that
\be
\label{60}
\sum_i p(A_i|X_j) = 1 \; .
\ee

Then we immediately come to the equality
$$
\sum_i q(A_i X) = 0 \; ,
$$
which is just a particular case of the general condition (\ref{46}).

This equality shows that, if at least one of the terms is non-zero, some
of the interference terms are necessarily negative and some are
necessarily positive. Therefore, some of the probabilities are depressed,
while others are enhanced. This alternation of the interference terms will
be shown below to be a pivotal feature providing a clear explanation of
the disjunction effect. It is worth emphasizing that the violation of the
sure-thing principle, resulting in the disjunction effect, will be shown
not to be due simply to the existence of interferences as such, but, more
precisely, to the {\it interference alternation}.

For instance, the depression of some probabilities can be associated with
uncertainty aversion, which makes less probable an action under uncertain
conditions. In contrast, the probability of other intentions, containing
less or no uncertainty, will be enhanced by positive interference terms.
This interference alternation is of crucial importance for the correct
description of decision making, without which the known paradoxes cannot
be explained.

\subsection{Less is More}

The title of this subsection is taken from a poem of the nineteenth
century English poet Robert Browning \cite{89bis}.

In the present context, this expression means that sometimes
excessive information is not merely difficult to get, but can even be
harmful, resulting in wrong decisions. It often happens that decisions,
based on smaller amount of information, are better than those based on
larger amount of information. This may happen because, with increasing
the amount of information, the choice between alternatives can become more
complicated as a result of which uncertainty grows. Increasing complexity
often increases uncertainty.

To describe the ``less is more'' phenomenon in mathematical
language, let us consider a prospect $\pi^*_k$ that is optimal under
a fixed information set $X_k$, with the probability \be \label{111}
p\left ( \pi^*_k \right ) = p_0\left ( \pi^*_k \right ) + q\left (
\pi^*_k \right ) \; . \ee Suppose, we increase the amount of
information by going to the information set $X_{k+1}$, such that
$X_k \in X_{k+1}$, and obtain the related optimal prospect
$\pi^*_{k+1}$, with the probability \be \label{112} p\left (
\pi^*_{k+1} \right ) = p_0\left ( \pi^*_{k+1} \right ) + q\left (
\pi^*_{k+1} \right ) \; . \ee

Assume that the utilities of these two prospects are the same,
\be
\label{113}
p_0\left ( \pi^*_{k+1} \right ) =
p_0\left ( \pi^*_k \right )  \; ,
\ee
while the uncertainty in the decision making process increases, so that
the attraction factor decreases,
\be
\label{114}
q\left ( \pi^*_{k+1} \right ) \; < \;
q\left ( \pi^*_k \right )  \; .
\ee

Then, the relation between the prospect probabilities
\be
\label{115}
p\left ( \pi^*_k \right ) -
p\left ( \pi^*_{k+1} \right ) = q\left ( \pi^*_k \right ) -
q\left ( \pi^*_{k+1} \right ) \; > \; 0 \;
\ee
tells us that the decision process leading to choosing prospect $\pi^*_k$
is clearer than for prospect $\pi^*_{k+1}$, because the larger
value of the corresponding probability makes the signal stronger
for the decision maker, resulting in a larger frequency of choices $\pi^*_k$.
As the information set is increased, in the presence of many alternatives,
the preferred prospect becomes less clearly defined as the top choice. As
a consequence, a lack of efficiency, a growing indeterminacy and
ultimately the freezing of the decision process can ensue.

When dealing with complex nonlinear problems, excessive information can
lead to incorrect conclusions because of the extreme sensitivity of
nonlinear problems to minor details. As simple examples, when excessive
information can be harmful, we may mention the following typical cases
from physics.

\vskip 2mm
{\bf Example 1}. How to describe the state of air in a room? The
unreasonable decision would be to analyze the motion of all molecules in
the room, specifying all their interactions, positions and velocities. Such
a decision would lead to not merely extremely overcomplicated calculations,
but even can result in incorrect conclusions. The reasonable decision is to
characterize the state of the air by defining the room temperature, volume,
and atmospheric pressure.

\vskip 2mm
{\bf Example 2}. How to characterize the water flow in a river? A silly
decision would be to consider the motion of all water molecules in the
river describing their locations, velocities, interactions, and so on.
Contrary to this, a clever decision is to use the hydrodynamic equations.

\vskip 2mm
{\bf Example 3}. How to describe a large social system? Again, the
unreasonable decision would be to collect all possible information on
each member of the society. Then, being overloaded by senseless information,
one would be lost in secondary details, being unable to make any clever
conclusion. Instead of this, it is often (though may be not always)
sufficient to consider the society composed of typical (or ``representative'')
agents.

\section{Disjunction Effect}

The disjunction effect was first specified by Savage \cite{58} as a violation
of the ``sure-thing principle'', which can be formulated as follows:
{\it If the alternative $A$ is preferred to the alternative $B$, when an
event $X_1$ occurs, and it is also preferred to $B$, when an event $X_2$
occurs, then $A$ should be preferred to $B$, when it is not known which
of the events, either $X_1$ or $X_2$, has occurred}.

\subsection{Sure-Thing Principle}

Let us now show how the sure-thing principle arises in
classical probability theory.

Let us consider a field of events $\{ A,B,X_j|j=1,2,\ldots\}$
equipped with the classical probability measures \cite{90}. We
denote the classical probability of an event $A$ by the capital
letter $P(A)$ in order to distinguish it from the probability $p(A)$
defined in the previous sections by means of quantum rules. We shall
denote, as usual, the conditional probability of $A$ under the
knowledge of $X$ by $P(A|X)$ and the joint probability of $A$ and
$X$, by $P(AX)$. We assume that at least one of the events $X_j$
from the set $\{ X_j\}$ certainly happens and that the $X_i$ are
mutually exclusive and exhaustive, which implies that \be \label{61}
\sum_j P(X_j) = 1 \; . \ee The probability of $A$, when $X_j$ is not
specified, that is, when at least one of $X_j$ happens, is denoted
by $P(AX)$, where $X = \bigcup_j X_j$. The same notations are
applied to $B$. Following the common wisdom, we understand the
statement ``$A$ is preferred to $B$" as meaning that $P(AX)>P(BX)$.
Then the following theorem is valid.

\vskip 3mm

{\bf Theorem 2}: {\it If for all $j=1,2,\ldots$, one has \be
\label{62} P(A|X_j) > P(B|X_j) \; , \ee then} \be \label{63} P(AX) >
P(BX) \; . \ee

{\it Proof}: It is straightforward that, under $X = \bigcup_j X_j$,
one has \be \label{64} P(AX) = \sum_j P(AX_j) = \sum_j
P(A|X_j)P(X_j) \ee and \be \label{65} P(BX) = \sum_j P(BX_j) =
\sum_j P(B|X_j)P(X_j) \; . \ee From Equations (\ref{64}) and
(\ref{65}), under assumption (\ref{62}), inequality (\ref{63})
follows immediately.

\vskip 2mm

The above proposition is a theorem of classical probability theory.
Savage \cite{58} proposed to use it as a normative statement on how human
beings make consistent decisions under uncertainty. As such, it is no
more a theorem but a testable assumption about human behavior. In other
words, empirical tests showing that humans fail to obey the sure-thing
principle must be interpreted as a failure of humans to abide to all
the rules of classical probability theory.

\subsection{Disjunction-Effect Examples}

Thus, according to standard classical probability theory which is
held by most statisticians as the only rigorous mathematical
description of risks, and therefore as the normative guideline
describing rational human decision making, the sure-thing principle
should be always verified in empirical tests involving real human
beings. However, numerous violations of this principle have been
investigated empirically \cite{58,91,92,93,94}. In order
to be more specific, let us briefly outline some examples of the
violation of the sure-thing principle, referred to as the
disjunction effect.

\vskip 2mm

(i) {\it To gamble or not to gamble}?

A typical setup for illustrating the disjunction effect is a
two-step gamble \cite{91}. Suppose that a group of people accepted a
gamble, in which the player can either win ($X_1$) or lose ($X_2$).
After one gamble, they are invited to gamble a second time, being
free to either accept the second gamble ($A$) or to refuse it ($B$).
Experiments by Tversky and Shafir \cite{91} showed that the majority
of people accept the second gamble when they know the result of the
first one, in any case, whether they won or lost in the previous
gamble. In the language of conditional probability theory, this
translates into the fact that people act as if $P(A|X_1)$ is larger
than $P(B|X_1)$ and $P(A|X_2)$ is larger than $P(B|X_2)$ as in
Equation (\ref{62}). At the same time, it turns out that the
majority refuses to gamble the second time when the outcome of the
first gamble is not known. The second empirical fact implies that
people act as if $P(BX)$ overweighs $P(AX)$, in blatant
contradiction with inequality (\ref{63}), which should hold
according to the theorem resulting from (\ref{62}). Thus,
a majority accepted the second gamble after having won or lost in
the first gamble, but only a minority accepted the second gamble
when the outcome of the first gamble was unknown to them. This
provides an unambiguous violation of the Savage sure-thing
principle.

\vskip 2mm

(ii) {\it To buy or not to buy}?

Another example, studied by Tversky and Shafir \cite{91}, had to do
with a group of students who reported their preferences about buying
a non-refundable vacation, following a tough university test. They
could pass the exam ($X_1$) or fail ($X_2$). The students had to
decide whether they would go on vacation ($A$) or abstain ($B$). It
turned out that the majority of students purchased the vacation when
they passed the exam as well as when they had failed, so that
condition (\ref{62}) was valid. However, only a minority of
participants purchased the vacation when they did not know the
results of the examination. Hence, inequality (\ref{63}) was
violated, demonstrating again the disjunction effect.

\vskip 2mm

(iii) {\it To sell or not to sell}?

The stock market example, analyzed by Shafir and Tversky \cite{95}, is a
particularly telling one, involving a deliberation taking into account
a future event, and not a past one as in the two previous cases. Suppose
we consider the USA presidential election, when either a Republican
wins ($X_1$) or a Democrat wins ($X_2$). On the eve of the election,
market players can either sell certain stocks from their portfolio ($A$)
or hold them ($B$). It is known that a majority of people would be inclined
to sell their stocks, if they would know who wins, regardless of whether
the Republican or Democrat candidate wins the upcoming election. This is
because people expect the market to fall after the elections. Hence,
condition (\ref{62}) is again valid. At the same time, a great many
people do not sell their stocks before knowing who really won the
election, thus contradicting the sure-thing principle and the inequality
(\ref{63}). Thus, investors could have sold their stocks before the
election at a higher price but, obeying the disjunction effect, they
were waiting until after the election, thereby selling at a lower price
after stocks have fallen. Many market analysts believe that this is
precisely what happened after the 1988 presidential election, when George
Bush defeated Michael Dukakis.

\vskip 2mm

There are plenty of other more or less complicated examples of the
disjunction effect \cite{58,91,92,93,95,96,97}. The
common necessary conditions for the disjunction effect to arise are
as follows. First, there should be several events, each
characterized by several alternatives, as in the two-step gambles.
Second, there should necessarily exist some uncertainty, whether
with respect to the past, as in the examples (i) and (ii), or with
respect to the future, as in the example (iii).

Several ways of interpreting the disjunction effect have been
analyzed. Here, we do not discuss the interpretations based on the
existence of some biases, such as the gender bias, or the bias
invoking the notion of decision complexity, which have already been
convincingly ruled out \cite{92,98}. We describe the reason-based
explanation which appears to enjoy a wide-spread following and
discuss its limits before turning to the view point offered by QDT.

\subsection{Reason-Based Analysis}

The dominant approach for explaining the disjunction effect is the
reason-based analysis of decision making \cite{91,92,95,97,99}. This
approach explains choice in terms of the balance between reasoning
for and against the various alternatives. The basic intuition is
that when outcomes are known, a decision maker may easily come up
with a definitive reason for choosing an option. However, in the
case of uncertainty, when the outcomes are not known, people may
lack a clear reason for choosing an option and consequently they
abstain and make an irrational choice.

From our perspective, the weakness of the reason-based analysis is
that the notion of ``reason'' is too vague and subjective. Reasons
are not only impossible to quantify, but it is difficult, if
possible at all, to give a qualitative definition of what they are.
Consider example (i) ``to gamble or not to gamble?''  Suppose you
have already won at the first step. Then, you can rationalize that
gambling a second time is not very risky: if you now loose, this
loss will be balanced by the first win on which you were not
counting anyway, so that you may actually treat it differently from
the rest of your wealth, according to the so-called ``mental
accounting'' effect; and if you win again, your profit will be
doubled. Thus, you have a ``reason'' to justify the attractiveness
of the second gamble. But, it seems equally justified to consider
the alternative ``reason'': if you have won once, winning the second
time may seem less probable (the so-called gambler's fallacy), and
if you loose, you will keep nothing of your previous gain. This line
of reasoning justifies to keep what you already got and to forgo the
second gamble.

Suppose now you have lost in the first gamble and know it. A first
reasoning would be that the second gamble offers a possibility of getting
out of the loss, which provides a reason for accepting the second gamble.
However, you may also think that the win is not guaranteed, and your
situation could actually worsen, if you loose again. Therefore, this makes
it more reasonable not to risk so much and to refrain from the new gamble.

Consider now the situation where you are kept ignorant of whether you have
won or lost in the first gamble. Then, you may think that there is no
reason and therefore no motivation for accepting the second gamble, which
is the standard reason-based explanation. But, one could argue that it
would be even more logical if you would think as follows:  Okay, I do not
know what has happened in the first gamble. So, why should I care about it?
Why don't I try again my luck? Certainly, there is a clear reason for
gambling that could propagate the drive to gamble a second time.

This discussion is not pretending to demonstrate anything other than
that the reason-based explanation is purely ad-hoc, with no real
explanatory power; it can be considered in a sense as a reformulation of
the disjunction fallacy. It is possible to multiply the number of examples
demonstrating the existence of quite ``reasonable'' justifications for
doing something as well as a reason for just doing the opposite. It seems
to us that the notion of ``reason" is not well defined and one can always
invent in this way a justification for anything. Thus, we propose that the
disjunction effect has no direct relation to reasoning. In the following
section, we suggest another explanation of this effect based on QDT,
specifically the interference between the two uncertain outcomes
resulting from an aversion to uncertainty (uncertainty-aversion principle),
which provides a {\it quantitative} testable prediction.

\subsection{Quantitative Analysis Within Quantum Decision Theory}

The disjunction effect, described above, finds a natural explanation in
the frame of the Quantum Decision Theory, as is shown below.

\vskip 3mm
{\bf Application to Disjunction-Effect Examples}

\vskip 2mm The possibility of connecting the violation of the
sure-thing principle with the occurrence of interference has been
mentioned in several articles (see, e.g., \cite{89}). But all these
attempts were ad hoc assumptions not based on a self-consistent
theory. Our explanation of the disjunction effect differs from these
attempts in several aspects. First, we consider the disjunction
effect as just one of several possible effects in the frame of the
{\it general theory}. The explanation is based on the theorem of
{\it interference alternation}, which has never been mentioned, but
without which no explanation can be complete and self-consistent. We
stress the importance of the {\it uncertainty-aversion principle}.
Also, we offer a {\it quantitative estimate} for the effect, which
is principally new.

Let us discuss the two first examples illustrating the disjunction
effect, in which the prospect consists of two intentions with two
representations each. One intention ``to decide about an action''
has the representations ``to act'' ($A$) and ``not to act'' ($B$).
The second intention ``to know the results" (or ``to have
information'') has also two representations. One ($X_1$) can be
termed ``to learn about the win'' (gamble won, exam passed), the
other ($X_2$) can be called ``to learn about the loss'' (gamble
lost, exam failed).  With the numbers of these representations
$M_1=2$ and $M_2=2$, the dimension of mind, given in Definition 7,
is $dim\cM = M_1 M_2=4$.

For the considered cases, the general set of Equations~(\ref{56})
reduces to two equations
$$
p(AX) = p(AX_1) + p(AX_2) + q(AX) \; ,
$$
\be
\label{66}
p(BX) = p(BX_1) + p(BX_2) + q(BX) \; ,
\ee
in which again $X = \bigcup_j X_j$ and the interference terms are the
attraction factors
$$
q(AX) = 2\vp(AX) \; \sqrt{p(AX_1)\; p(AX_2) } \; ,
$$
\be \label{67} q(BX) = 2\vp(BX) \; \sqrt{p(BX_1)\; p(BX_2) } \; .
\ee Of course, Equations (\ref{66}) and (\ref{67}) could be
postulated, but then it would not be clear where they come from. In
QDT, these equations appear naturally. Here $\vp(AX)$ and $\vp(BX)$
are the uncertainty factors defined in (\ref{53}). The
normalizations (\ref{58}) and (\ref{59}) become \be \label{68} p(AX)
+ p(BX) = 1 \; , \qquad p(X_1) + p(X_2) = 1\; . \ee The
normalization condition (\ref{60}) gives \be \label{69} p(A|X_1) +
p(B|X_1) = 1 \; , \qquad p(A|X_2) + p(B|X_2) = 1 \; . \ee The
uncertainty factors can be rewritten as \be \label{70} \vp(AX) =
\frac{q(AX)}{2\sqrt{p(AX_1)p(AX_2)} } \; , \qquad \vp(BX) =
\frac{q(BX)}{2\sqrt{p(BX_1)p(BX_2)} } \; , \ee with the interference
terms being \be \label{71} q(AX) = p(AX) - p(AX_1) - p(AX_2) \; ,
\qquad q(BX) = p(BX) - p(BX_1) - p(BX_2) \; . \ee

The principal point is the condition of {\it interference alternation}
(\ref{46}), which now reads
\be
\label{72}
q(AX) + q(BX) = 0 \; .
\ee
Without this condition (\ref{72}), the system of equations for the
probabilities would be incomplete, and the disjunction effect could not
be explained.

In the goal of explaining the disjunction effect, it is not sufficient to
merely state that some type of interference is present. It is necessary to
determine (quantitatively if possible) why the probability of acting is
suppressed, while that of remaining passive is enhanced. Our aim is to
evaluate the expected size and signs of the interference terms $q(AX)$
(for acting under uncertainty) and $q(BX)$ (for remaining inactive
under uncertainty). Obviously, it is an illusion to search for a universal
value that everybody will use. Different experiments with different people
have indeed demonstrated a significant heterogeneity among people, so that,
in the language of QDT, this means that the values of the interference terms
can fluctuate from individual to individual. A general statement should here
refer to the behavior of a sufficiently large ensemble of people, allowing
us to map the observed frequentist distribution of decisions to the
predicted QDT probabilities.

\vskip 3mm
{\bf Attraction Factors as Interference Terms }

\vskip 2mm
The interference terms (\ref{67}) can be rewritten as
$$
q(AX) = 2\vp(AX) \; \sqrt{p(A|X_1) p(X_1) p(A|X_2) p(X_2)} \; ,
$$
\be \label{73} q(BX) = 2\vp(BX) \; \sqrt{p(B|X_1) p(X_1) p(B|X_2)
p(X_2)} \; . \ee The interference-alternation theorem (Theorem 1),
which leads to Equation (\ref{72}), implies that \be |q(AX)| =
|q(BX)| \; , \label{74} \ee and \be {\rm sign}[\vp(AX)] = - {\rm
sign}[\vp(BX)]~. \label{75} \ee Hence, in the case where $p(A|X_j) >
p(B|X_j)$, which is characteristic of the examples illustrating the
disjunction effect, one must have the uncertainty factors which
exhibit the opposite property, $|\vp(AX)|< |\vp(BX)|$, so as to
compensate the former inequality to ensure the validity of the
equality (\ref{74}) for the absolute values of the interference
terms. The next step is to determine the sign of $\vp(AX)$ (and thus
of $\vp(BX)$) from (\ref{75}) and their typical amplitudes
$|\vp(AX)|$ and $|\vp(BX)|$.

\vskip 3mm
{\bf Signs of Uncertainty Factors}.

\vskip 2mm A fundamental well-documented characteristic of human
beings is their aversion to uncertainty, {\it i.e.}, the preference
for known risks over unknown risks \cite{100}. As a consequence, the
propensity/utility (and therefore the probability) to act under
larger uncertainty is smaller than under smaller uncertainty.
Mechanically, this implies that it is possible to specify the sign
of the uncertainty factors, yielding \be {\rm sign}[\vp(AX)] = -
{\rm sign}[\vp(BX)] <0~, \label{76} \ee since $A$ (respectively $B$)
refers to acting (respectively, to remain inactive).

\vskip 3mm
{\bf Amplitudes of Uncertainty Factors}.

\vskip 2mm As a consequence of Equation (\ref{76}) and also of their
mathematical definition (\ref{53}), the uncertainty factors vary in
the intervals \be \label{77} -1 \leq \vp(AX) \leq 0 \; , \qquad 0
\leq \vp(BX) \leq 1 \; . \ee Without any other information, the
simplest prior is to assume a uniform distribution of the
uncertainty factors in each interval, so that their expected values
are respectively \be \label{78} \overline\vp(AX) = -\; \frac{1}{2}
\; , \qquad \overline\vp(BX) = \frac{1}{2} \; . \ee Choosing in that
way the average values of the uncertainty factors is equivalent to
using a representative agent, while the general approach is fully
taking into account a pre-existing heterogeneity. That is, the
values (\ref{78}) should be treated as estimates for the expected
uncertainty factors, corresponding to these factors averaged with
the uniform distribution over the large number of agents.

\vskip 3mm
{\bf Interference-Quarter Law}

\vskip 3mm To complete the calculation of $q(AX)$ and of $q(BX)$
given by Equations (\ref{73}), we also assume the non-informative
uniform prior for all probabilities appearing below the
square-roots, so that their expected values are all $1/2$ since they
vary between $0$ and $1$. Using these in Equation~(\ref{73}) results
in the interference-quarter law \be \label{79} \overline q(AX) =
-0.25 \; , \qquad \overline q(BX) = 0.25 \; , \ee valid for the
four-dimensional mind composed of two intentions with two
representations each.

As a consequence, the probabilities for acting or for remaining
inactive under uncertainty, given by Equations (\ref{66}), can be
evaluated as
$$
p(AX) = p(AX_1) + p(AX_2) - 0.25 \; ,
$$
\be
\label{80}
p(BX) = p(BX_1) + p(BX_2) + 0.25 \; .
\ee
The influence of intention interference, in the presence of uncertainty,
on the decision making process at the basis of the disjunction effect can
thus be estimated a priori. The sign of the effect is controlled by the
aversion to uncertainty exhibited by people (uncertainty-aversion
principle). The amplitude of the effect can be estimated, as shown above,
from simple priors applied to the mathematical structure of the QDT
formulation.

\vskip 3mm
{\bf Uncertainty-Aversion Principle}

\vskip 3mm
The above calculation implies that the disjunction effect can be
interpreted as essentially an emotional reaction associated with the
{\it aversion to uncertainty}. An analogy can make the point: it is
widely recognized that uncertainty frightens living beings, whether
humans or animals. It is also well documented that fear paralyzes, as in
the cartoon of the ``rabbit syndrome,'' when a rabbit stays immobile in
front of an approaching boa instead of running away. There are many
circumstantial evidences that uncertainty may frighten people as a boa
frightens rabbits. Being afraid of uncertainty, a majority of human
beings may be hindered to act. In the presence of uncertainty, they do
not want to act, so that they refuse the second gamble, as in example (i),
or forgo the purchase of a vacation, as in example (ii), or refrain
from selling stocks, as in example (iii). Our analysis suggests that it
is the aversion to uncertainty that paralyzes people and causes the
disjunction effect.

It has been reported that, if people, when confronting uncertainty
paralyzing them against acting, are presented with a detailed
explanation of the possible outcomes, they then may change their
mind and decide to act, thus reducing the disjunction effect
\cite{91,92}. Thus, by encouraging people to think by providing them
additional explanations, it is possible to influence their minds. In
such a case, reasoning plays the role of a kind of therapeutic
treatment decreasing the aversion to uncertainty. This line of
reasoning suggests that it should be possible to decrease the
aversion to uncertainty by other means, perhaps by distracting them
or by taking food, drink or drug injections. This provides the
possibility to test for the dependence of the strength of the
disjunction effect with respect to various parameters which may
modulate the aversion response of individuals to uncertainty.

We should stress that our explanation departs fundamentally from the
standard reason-based rationalization of the disjunction effect summarized
above. Rather than using what we perceive is an hoc explanation, we anchor
the disjunction effect on the very fundamental characteristic of living
beings, that of the aversion to uncertainty. This allows us to construct
a robust and parsimonious explanation. But this explanation arises only
within QDT, because the latter allows us to account for the complex
emotional, often subconscious, feelings as well as many unknown states of
nature that underlie decision making. Such unknown states, analogous to
hidden variables in quantum mechanics, are taken into account by the
formalism of QDT through the interference alternation effect, capturing
mental processes by means of quantum-theory techniques.

It is appropriate here to remind once more that it was Bohr who
advocated throughout all his life the idea that mental processes do
bear close analogies with quantum processes (see, e.g.,
\cite{36,37,38,39}). Since interference is one of the most striking
characteristic features of quantum processes, the analogy suggests
that it should also arise in mental processes as well. The existence
of interference in decision making disturbs the classical additivity
of probabilities. Indeed, we take as an evidence of this the
nonadditivity of probabilities in psychology which has been
repeatedly observed \cite{101,102,103}, although it has not been
connected with interference.

\vskip 3mm
{\bf Numerical Analysis of Disjunction Effect}

\vskip 2mm
In the frame of QDT, it is possible, not merely to connect the existence
of the disjunction effect with interference, but to give {\it quantitative
predictions}. Below, this is illustrated by the numerical explanation of
the examples described above.

(i) {\it To gamble or not to gamble}?

Let us turn to the example of gambling. The statistics reported by Tversky
and Shafir \cite{91} are
$$
p(A|X_1)=0.69 \; , \qquad p(A|X_2)=0.59 \; , \qquad p(AX)=0.36 \; .
$$
Then Equations (\ref{68}) and (\ref{69}) give
$$
p(B|X_1)=0.31 \; , \qquad p(B|X_2)=0.41 \; , \qquad p(BX)=0.64 \; .
$$
Recall that the disjunction effect here is the violation of the sure-thing
principle, so that, although $p(A|X_j)>p(B|X_j)$ for $j=1,2$, one observes
nevertheless that $p(AX)<p(BX)$. In the experiment reported by Tversky and
Shafir \cite{91}, the probabilities for winning or for losing were identical:
$p(X_1)=p(X_2)=0.5$. Then, using relation (\ref{57}), we obtain
$$
p(AX_1) =0.345\; , \qquad p(AX_2)=0.295\; , \qquad
p(BX_1) =0.155\; , \qquad p(BX_2)=0.205 \; .
$$
For the interference terms, we find
\be
q(AX) = -0.28\; , \qquad q(BX) = 0.28 \; .
\label{81}
\ee
The uncertainty factors (\ref{70}) are therefore
$$
\vp(AX) = -0.439 \; , \qquad \vp(BX) = 0.785 \; .
$$
They are of opposite sign, in agreement with condition (\ref{75}).
The probability $p(AX)$ of gambling under uncertainty is suppressed by
the negative interference term $q(AX)<0$. Reciprocally, the probability
$p(BX)$ of not gambling under uncertainty is enhanced by the positive
interference term $q(BX)>0$. This results in the disjunction effect,
when $p(AX)<p(BX)$.

It is important to stress that the observed amplitudes in (\ref{81})
are close to the interference-quarter law (\ref{79}). Actually, within
the experimental accuracy with a statistical error about $20\%$, the found
interference terms cannot be distinguished from the value 0.25. Thus, even
not knowing the results of the considered experiment, we are able to
{\it quantitatively} predict the strength of the disjunction effect.

\vskip 4mm
(ii) {\it To buy or not to buy}?

For the second example of the disjunction effect, the data, taken
from \cite{91}, read
$$p(A|X_1)=0.54 \; , \qquad p(A|X_2)=0.57 \; , \qquad p(AX) = 0.32 \; .
$$
Following the same procedure as above, we get
$$
p(B|X_1)=0.46 \; , \qquad p(B|X_2)=0.43 \; , \qquad p(BX) = 0.68 \; .
$$
Given again that the two alternative outcomes are equiprobable,
$p(X_1)=p(X_2)=0.5$, we find
$$
p(AX_1) =0.270\; , \qquad p(AX_2)=0.285\; , \qquad
p(BX_1) =0.230\; , \qquad p(BX_2)=0.215 \; .
$$
For the interference terms, we obtain
\be
q(AX) = - 0.235 \; , \qquad q(BX) = 0.235 \; .
\label{82}
\ee
The uncertainty factors are
$$
\vp(AX)=-0.424 \; , \qquad \vp(BX)=0.528 \; .
$$
Again, the values obtained in (\ref{82}) are close to our predicted
interference-quarter law (\ref{79}). More precisely, these values
are actually undistinguished from 0.25 within the statistical error
$20\%$, typical of the discussed experiments.

Because of the uncertainty aversion, the probability $p(AX)$ of
purchasing a vacation is suppressed by the negative interference
term $q(AX)<0$. At the same time, the probability $p(BX)$ of not
buying a vacation under uncertainty is enhanced by the positive
interference term $q(BX)>0$. This alternation of interferences
causes the disjunction effect resulting in $p(AX)<p(BX)$. It is
necessary to stress it again that without this interference
alternation no explanation of the disjunction effect is possible
in principle.

In the same way, our approach can be applied to any other situation
related to the disjunction effect associated with the violation of the
sure-thing principle. We now turn to another deviation from rational
decision making, known under the name of the conjunction fallacy.

\section{Conjunction Fallacy}

The conjunction fallacy constitutes another example revealing that
intuitive estimates of probability by human beings do not conform to
the standard probability calculus. This effect was first studied by
Tversky and Kahneman \cite{104,105} and then discussed in many other
works (see, e.g., \cite{96,106,107,108,109}). Despite an
extensive debate and numerous attempts to interpret this effect,
there seems to be no consensus on the origin of the conjunction
fallacy \cite{109}.

Here, we show that this effect finds a natural explanation in QDT.
It is worth emphasizing that we do not invent a special scheme for
this particular effect, but we show that it is a natural consequence
of the general theory we have developed. In order to claim to
explain the conjunction fallacy in terms of an interference effect
in a quantum description of probabilities, it is necessary to derive
the quantitative values of the interference terms, amplitudes and
signs, as we have done above for the examples illustrating the
disjunction effect. This has never been done before. Our QDT
provides the necessary ingredients, in terms of the
uncertainty-aversion principle, the theorem on interference
alternations, and the interference-quarter law. Only the
establishment of these general laws can provide an explanation of
the conjunction fallacy, that can be taken as a positive step
towards validating QDT, according to the general methodology of
validating theories \cite{110}. Finally, in our comparison with
available experimental data, we analyze a series of experiments and
demonstrate that all their data substantiate the validity of the
general laws of the theory.

\subsection{Conjunction Rule}

Let us first briefly recall the conjunction rule of standard
probability theory. Let us consider an event $A$ that can occur
together with another one among several other events $X_j$, where
$j=1,2,\ldots$. The probability of an event, estimated within
classical probability theory, is again denoted with the capital
letter $P(A)$, to distinguish it from the probability $p(A)$ in our
quantum approach. According to standard probability theory
\cite{90}, one has \be \label{83} P(AX) = \sum_j P(AX_j) \; . \ee
Since all terms in the sum (\ref{83}) are positive, the conjunction
rule tells us that \be \label{84} P(AX) \geq P(AX_j) \qquad (\forall
~ j~). \ee That is, the probability for the occurrence of the
conjunction of two events is never larger than the probability for
the occurrence of a separate event.

\subsection{Conjunction Error}

Counterintuitively, humans rather systematically violate the conjunction
rule (\ref{84}), commonly making statements such that
\be
\label{85}
p(AX) < p(AX_j) \; ,
\ee
for some $j$, which is termed the {\it conjunction fallacy} (Tversky and
Kahneman \cite{104,105}). The difference
\be
\label{86}
\ep (AX_j) \equiv p(AX_j) - p(AX)
\ee
is called the {\it conjunction error}, which is positive under conditions
in which the conjunction fallacy is observed.

A typical situation is when people judge about a person, who can possess
a characteristic $A$ and also some other characteristics $X_j$. This, e.g.,
can be ``possessing a trait'' or ``not having the trait'', since
not having a trait is also a characteristic.  The often-cited example
of Tversky and Kahneman \cite{104} is as follows: ``Linda is 31 years old,
single, outspoken, and very bright. She majored in philosophy. As a
student, she was deeply concerned with issues of discrimination and social
justice, and also participated in anti-nuclear demonstrations. Which is
more likely? (i) Linda is a bank teller; (ii) Linda is a bank teller and
is active in the feminist movement.''  Most people answer (ii) which is an
example of the conjunction fallacy (\ref{85}).

Numerous other examples of the fallacy are described in the
literature \cite{96,105,106,107,108,109}. It is important to stress
that this fallacy has been reliably and repeatedly documented, that
it cannot be explained by the ambiguity of the word ``likely'' used
in the formulation of the question, and that it appears to involve a
failure to coordinate the logical structure of events in the
presence of chance \cite{109}. The conjunction fallacy cannot be
explained by prospect theory \cite{111} and also remains when
different bracketing effects \cite{112,113,114,115,116} are
taken into account.

\subsection{Conjunction Interference}

Within QDT, the conjunction fallacy finds a simple and natural
explanation. Let us consider a typical situation of the fallacy, when
one judges a person who may have a characteristic $A$, treated as primary,
and who may also possess, or not possess, another characteristic, labeled
as secondary. Generally, the person could also be an object, a fact, or
anything else, which could combine several features. Translating this
situation to the mathematical language of QDT, we see that it involves
two intentions. One intention, with just one representation, is ``to decide
whether the object has the feature $A$.'' The second intention ``to decide
about the secondary feature" has two representations, when one decides
whether ``the object has the special characteristic" $(X_1)$ or ``the
object does not have this characteristic" $(X_2)$.

For these definitions, and following the general scheme, we have
\be
\label{87}
p(AX) = p(AX_1) + p(AX_2) + q(AX) =
p(A|X_1)p(X_1) + p(A|X_2)p(X_2) + q(AX) \; .
\ee
This is a typical situation where a decision is taken under uncertainty.
The uncertainty-aversion principle requires that the interference term
$q(AX)$ should be negative ($q(AX)<0$). Indeed, this reflects
that the probability for a human to act under larger uncertainty is smaller
than under smaller uncertainty, in line with definition (\ref{70}) and
condition (\ref{76}).

Taking the perspective of the representation $X_1$, definition
(\ref{86}) together with Equations (\ref{87}) imply that the
conjunction error reads \be \label{88} \ep(AX_1) = | q(AX)| -
p(AX_2) \; . \ee The condition for the conjunction fallacy to occur
is that the error (\ref{88}) be positive, which requires that the
interference term be sufficiently large, such that the {\it
conjunction-fallacy condition} \be \label{89} | q(AX)| > p(AX_2) \ee
be satisfied.

The QDT thus predicts that a person will make a decision exhibiting
the conjunction fallacy when (i) uncertainty is present and (ii) the
interference term, which is negative by the uncertainty-aversion
principle, has a sufficiently large amplitude, according to
condition (\ref{89}).

\subsection{Comparison with Experiments}

For a quantitative analysis, we take the data from Shafir {\it et
al}. \cite{96}, who present one of the most carefully accomplished
and thoroughly discussed set of experiments. Shafir {\it et al}.
questioned large groups of students in the following way. The
students were provided with booklets each containing a brief
description of a person. It was stated that the described person
could have a primary characteristic $(A)$ and could have
additionally a second characteristic $(X_1)$, or could be free of
this second characteristic $(X_2)$.

In total, there were 28 experiments separated into two groups
according to the conjunctive category of the studied
characteristics. In 14 cases, the features $A$ and $X_1$ were
compatible with each other, and in the other 14 cases, they were
incompatible. The characteristics were treated as compatible, when
they were felt as closely related according to some traditional
wisdom, for instance, ``woman teacher'' $(A)$ and ``feminist''
$(X_1)$. Another example of compatible features is ``chess player''
$(A)$ and ``professor'' $(X_1)$. Those characteristics that were not
related by direct logical connections were considered as
incompatible, such as ``bird watcher'' $(A)$ and ``truck driver''
$(X_1)$ or ``bicycle racer'' $(A)$ and ``nurse'' $(X_1)$.

In each of the 28 experiments, the students were asked to evaluate both
the typicality and the probability of $A$ and $AX_1$. Since normal people
usually understand ``typicality'' just as a synonym of probability, and
vice versa, the prediction on typicality were equivalent to estimates of
probabilities. This amounts to considering only how the students estimated
the probability $p(AX)$, with $X = X_1 + X_2$, that the considered person
possesses the stated primary feature and the probability $p(AX_1)$ that
the person has both characteristics $A$ and $X_1$.

An important quality of the experiments by Shafir {\it et al}.
\cite{96} lies in the large number of tests which were performed.
Indeed, a given particular experiment is prone to exhibit a
significant amount of variability, randomness or ``noise''. Not only
the interrogated subjects exhibited significant idiosyncratic
differences, with diverse abilities, logic, and experience, but, in
addition, the questions were quite heterogeneous. Even the
separation of characteristics into two categories of compatible and
incompatible pairs is to a great extent arbitrary. Consequently, no
one particular case provides a sufficiently clear-cut conclusion on
the existence or absence of the conjunction effect. It is only by
realizing a large number of interrogations, with a variety of
different questions, and by then averaging the results, that it is
possible to make justified conclusions on whether or not the
conjunction fallacy exists. The set of experiments performed by
Shafir {\it et al}. \cite{96} well satisfies these requirements.

For the set of compatible pairs of characteristics, it turned out
that the average probabilities were $p(AX)=0.537$ and
$p(AX_1)=0.567$, with statistical errors of $20\%$. Hence, within
this accuracy, $p(AX)$ and $p(AX_1)$ coincide and no conjunction
fallacy arises for compatible characteristics. From the view point
of QDT, this is easily interpreted as due to the lack of
uncertainty: since the features $A$ and $X_1$ are similar to each
other, one almost certainly yielding the other, there is no
uncertainty in deciding, hence, no interference, and, consequently,
no conjunction fallacy.

However, for the case of incompatible pairs of characteristics, the
situation was found to be drastically different. To analyze the
related set of experiments, we follow the general scheme of the
previous subsection, using the same notations. We have the prospect
with two intentions, one intention is to evaluate a primary feature
$(A)$ of the object, and another intention is to decide whether, at
the same time, the object possesses a secondary feature $(X_1)$ or
does not possess it $(X_2)$. Taking the data for $p(X_j)$ and
$p(AX_1)$ from Shafir {\it et al}. \cite{96}, we calculate $q(AX)$
for each case separately and then average the results. In the
calculations, we take into account that the considered pairs of
characteristics are incompatible with each other. The simplest and
most natural mathematical embodiment of the property of
``incompatibility'' is to take the probabilities of possessing $A$,
under the condition of either having or not having $X_1$, as equal,
that is, $p(A|X_j)=0.5$. For such a case of incompatible pairs of
characteristics, Equation (\ref{87}) reduces to \be \label{90} p(AX)
= \frac{1}{2} + q(AX) \; . \ee The results, documenting the
existence of the interference terms underlying the conjunction
fallacy, are presented in Table 1, which gives the abbreviated names
for the object characteristics, whose detailed description can be
found in Shafir {\it et al}. \cite{96}.

\begin{table}[h!]
\caption{Conjunction fallacy and related interference terms caused
by the decision under uncertainty. The average interference term is
in good agreement with the interference-quarter law. The empirical
data are taken from Shafir {\it et al}. \cite{96}.}
\begin{center}
\begin{tabular}{|c|c|c|c|c|}\hline
  & characteristics & $p(AX)$ & $p(AX_1)$ & $q(AX)$ \\ \hline
$A$   & bank teller      &   0.241 & 0.401 & -0.259 \\
$X_1$ & feminist        &         &       &       \\ \hline
$A$   & bird watcher    & 0.173   & 0.274 & -0.327 \\
$X_1$ & truck driver    &         &       &       \\ \hline
$A$   & bicycle racer   & 0.160   & 0.226 & -0.340 \\
$X_1$ & nurse           &         &       &        \\ \hline
$A$   & drum player     & 0.266   & 0.367 & -0.234 \\
$X_1$ & professor       &         &       &        \\ \hline
$A$   & boxer           & 0.202   & 0.269 & -0.298 \\
$X_1$ & chef            &         &       &         \\ \hline
$A$   & volleyboller    & 0.194   & 0.282 & -0.306 \\
$X_1$ & engineer        &         &       &        \\ \hline
$A$   & librarian       & 0.152   & 0.377 & -0.348 \\
$X_1$ & aerobic trainer &         &       &        \\ \hline
$A$   & hair dresser    & 0.188   & 0.252 & -0.312  \\
$X_1$ & writer          &         &       &     \\ \hline
$A$   & floriculturist  & 0.310   & 0.471 & -0.190 \\
$X_1$ & state worker    &         &       &    \\ \hline
\end{tabular}
\end{center}
\end{table}

\begin{table}[h!]
\begin{center}
Table 1 {\it Cont}.\\
\begin{tabular}{|c|c|c|c|c|}\hline
$A$   & bus driver      & 0.172   & 0.314 & -0.328 \\
$X_1$ & painter         &         &       &    \\ \hline
$A$   & knitter         & 0.315   & 0.580 & -0.185 \\
$X_1$ & correspondent   &         &       &   \\ \hline
$A$   & construction worker & 0.131 & 0.249 & -0.369 \\
$X_1$ & labor-union president &    &      &    \\ \hline
$A$   & flute player    & 0.180   & 0.339 & -0.320 \\
$X_1$ & car mechanic    &         &       &    \\ \hline
$A$   & student         & 0.392   & 0.439 & -0.108 \\
$X_1$ & fashion-monger   &         &       &       \\ \hline
      & average         & 0.220   & 0.346 & -0.280 \\ \hline
\end{tabular}
\end{center}
\end{table}

The average values of the different reported probabilities are
$$
p(AX) = 0.22   \; , \qquad p(X_1)=0.692 \; , \qquad p(X_2)=0.308 \; ,
$$
\be p(AX_1)=0.346 \; ,  \qquad p(AX_2)=0.154. \label{91} \ee One can
observe that the interference terms fluctuate around a mean of
$-0.28$, with a standard deviation of $\pm 0.06$, that is \be
\label{92} \overline q(AX) = - 0.28\pm 0.06 \; . \ee There is a
clear evidence of the conjunction fallacy, with the conjunction
error (\ref{86}) being $\ep(AX_1)=0.126$.

QDT interprets the conjunction effect as due to the uncertainty
underlying the decision, which leads to the appearance of the
intention interferences. The interference of intentions is caused
by the hesitation whether, under the given primary feature $(A)$,
the object possesses the secondary feature $(X_1)$ or does not have
it $(X_2)$.  The term $\overline q(AX)$ is negative, reflecting
the effect of deciding under uncertainty (according to the
uncertainty-aversion principle). Quantitatively, we observe that
the amplitude $|\overline q(AX)|$ is in agreement with the QDT
interference-quarter law.

\subsection{Combined Conjunction and Disjunction Effects}

The QDT predicts that setups in which the conjunction fallacy occurs
should also be accompanied by the disjunction effect. To see this,
let us extend slightly the previous decision problem by allowing for
two representations of the first intention. Concretely, this means
that the intention, related to the decision about the primary
characteristic, has two representations: (i) ``decide about the
object or person having or not the primary considered feature''
$(A)$, and  (ii)  ``decide to abstain from deciding about this
feature'' $(B)$. This frames the problem in the context analyzed in
the previous section. The conjunction fallacy occurs when one
considers incompatible characteristics \cite{96,105}, such that the
probabilities of deciding of having a conjunction $(AX_j)$ or of not
guessing about it $(BX_j)$ are close to each other, so that one
can set \be \label{93} p(A|X_j) = p(B|X_j) \qquad
(\forall j) \; . \ee The theorem on interference alternation
(Theorem 1) implies that the interference term for being passive
under uncertainty is positive and we have \be \label{94} q(BX) = -
q(AX) > 0 \; . \ee Now, the probability $p(BX)$ of deciding not to
guess under uncertainty is governed by an equation similar to
Equation (\ref{87}). Combining this equation with (\ref{94}), we
obtain \be \label{95} p(BX) = p(AX) + 2 | q(AX) | \; , \ee which
shows that, despite equality (\ref{93}), the probability of being
passive is larger than the probability of acting under uncertainty.
This is nothing but a particular case of the disjunction effect.

This example shows that the conjunction fallacy is actually a sufficient
condition for the occurrence of the disjunction effect, both resulting
from the existence of interferences between probabilities under
uncertainty. The reverse does not hold: the disjunction effect does not
necessarily yield the conjunction fallacy, because the latter requires not
only the existence of interferences, but also that their amplitude should
be sufficiently large according to the conjunction-fallacy condition
(\ref{89}).

To our knowledge, experiments or situations when the disjunction and
conjunction effects are observed simultaneously have not been investigated.
The specific prediction coming from the QDT, that the disjunction effect
should be observable as soon as the conjunction effect is present,
provides a good test of QDT.

\section{Non-commutativity of Decisions}

It has been mentioned that subsequent decisions, in general, do not
commute with each other and that the non-commutativity is intimately
connected with the presence of interferences between intentions. As
is demonstrated in the previous sections, the phenomenon of
intention interference is a key and general phenomenon at the basis
of the disjunction effect and conjunction fallacy. Within QDT, we
expect it to be generically present in human decision making. We are
now in a position to present a rigorous proof that the phenomenon of
intention interference is also a crucial ingredient for
understanding the non-commutativity of successive decisions.

\subsection{Mathematical Formulation of Non-commutativity}

To describe in precise mathematical terms the property of
non-commutativity, let us consider the case of two intentions. We
denote one intention as $A$. And let the other intention $X \equiv
\bigcup_i X_i$, with $i=1,2,3,\ldots\}$ be composed of several
representations $X_i$, such that the intention $A$ can be certainly
realized under one of the intentions $X_i$ from the family $X$, that
is, \be \label{96} \sum_i p (X_i|A) = 1 \; . \ee Assume that the
joint probabilities are related to the conditional probabilities in
the standard way, such that \be \label{97} p(AX_i) \equiv
p(A|X_i)p(X_i) \; , \qquad  p(X_iA) \equiv p(X_i|A) p(AX) \; . \ee
For two intended actions $A$ and $X_i$ the following statement
holds, demonstrating the non-commutativity of these intended
actions.

\vskip 3mm

{\bf Theorem 3}: {\it For two intended actions, $A$ and $X=\bigcup_i X_i$,
satisfying conditions (\ref{96}) and (\ref{97}), the joint probability
$p(AX)$ equals $p(XA)$ if and only if there is no interference terms,
\be
\label{98}
p(AX) = p(XA) \; \; \leftrightarrow \; \; q(AX) = q(XA) \equiv 0 \; .
\ee
And, reciprocally, the intended actions $A$ and $X$ do not commute if and
only if the interference factors are nonzero,
\be
\label{99}
p(AX) \neq p(XA) \; , \qquad q(AX) \not\equiv 0 \; .
\ee
}

\vskip 2mm

{\it Proof}: By the general rules of QDT, we have
$$
p(XA) = \sum_i p(X_iA) + q(XA) \;  .
$$
Employing equations (97) gives
$$
p(XA) = \sum_i p(X_i|A) p(AX) + q(XA) \;  .
$$
Using normalization (96) yields
$$
p(XA) - p(AX) = q(XA) \;  .
$$
Interchanging here the actions $A$ and $X$ results in
$$
p(XA) - p(AX) = q(AX) \;  .
$$
The latter two equations prove the theorem.

\vskip 2mm

The non-commutativity of subsequent decisions is reminiscent of the
non-commutativity of subsequent measurements in quantum mechanics.
However, there is a principal difference between these phenomena. In
decision theory, the prospect states and the strategic state of mind
are the {\it internal states} of the same decision maker. In
contrast, in quantum mechanics, the measurement is accomplished by
an observer, or an apparatus, which are {\it external} to the
measured physical system. The analogy would be closer, if one could
imagine a physical system that attempts to measure some parts of
itself. Since standard quantum mechanical measurements do not
proceed like this, the mathematics of the non-commutativity of
subsequent decisions in decision theory and of subsequent
measurements in quantum theory are quite different.

\subsection{Meaning of Simultaneous Intended Actions}

As follows from the above theorem, when there are two intentions, say
$A$ and $B$, the joint probability $p(AB)$ is generally different from
$p(BA)$. Two intentions do not commute with each other, when at least one
of them is composite, consisting of several interfering representations,
or modes. The intentions commute, only when there is no mode
interference. For example, when the mind is one-dimensional or if there
is no uncertainty.

Since the order of intended actions is important, when writing $p(AB)$,
one has to keep in mind that the intention $B$ is to be realized earlier
than $A$. Even when talking about simultaneous intentions, it is implied
that the order $AB$ means the possible realization of $B$ infinitesimally
earlier than that of $A$. To be more precise, let us mark the intention $A$,
associated with time $t$, as $A_t$. Respectively, $B_t$ is the intention
$B$, associated with time $t$. Then the joint probability of these two
intentions, taken in the order $A_tB_t$, is defined as
\be
\label{100}
p(A_tB_t) \equiv \lim_{t'\ra t+0} \; p(A_{t'}B_t) \; .
\ee

Because of the non-commutativity of two intentions, the
corresponding decisions also do not commute. Two subsequent
decisions, even taken immediately one after another, and under the
same circumstances, in general, may lead to different outcomes just
as a result of the order of their realization.

\section{Entropy and Information Functional}

Quantum decision theory is developed above as a self-consistent
mathematical theory. But it remains to be shown how this general
theory could be reduced to classical decision theory as a particular
case. It is thus necessary to explain how QDT is connected to
classical decision theory based on the notion of expected utility.
For this purpose, we need to spell out the relation between the
utility factor (\ref{17}) and the classical expected utility. This
can be done by invoking conditional entropy maximization, which is
equivalent to the minimization of an information functional. The
method of conditional entropy maximization is widely used in
statistical physics yielding Gibbs ensembles \cite{117}.
The method of information minimization is in the basis of the
approach to constructing {\it representative ensembles}
\cite{118,119}, using which it is possible to obtain self-consistent
description of all, even quite complex, phenomena of statistical
physics.

The utility factors $p_0(\pi_j)$ in QDT play the role of classical
probabilities. Then the entropy can be defined through these utility
factors in the standard way as for any probabilities:
\be
\label{116}
S = - \sum_j p_0(\pi_j) \ln p_0(\pi_j)\; ,
\ee
where the summation is over all prospects $\pi_j$ pertaining to the given
prospect lattice $\cal L$ (see Definition 5). The normalization condition
(\ref{18}) is valid.

In classical decision theory, one deals with expected utilities
defined for the related lotteries \cite{57}. For each prospect
$\pi_j$, it is possible \cite{74} to put into correspondence a
lottery $L_j$. Therefore, the expected utility for a lottery $L_j$
can be denoted as depending on the prospect $\pi_j$ corresponding to
this lottery. So, we can write $U(\pi_j)$ for an expected utility
of a prospect $\pi_j$ related to the lottery $L_j$. For
concreteness, we assume that the expected utility is defined so that
it is non-negative: \be \label{117} U(\pi_j) \geq 0 \qquad (\pi_j
\in {\cal L}) \; , \ee which is always possible to achieve.

In addition to the normalization condition (\ref{18}), we have to impose
another condition related to the choice of expected utilities. To this end,
we shall use the notion of the likelihood ratio, known in testing
statistical theories \cite{120}. In classical decision theory, that lottery is
classified as optimal, which provides the maximal expected utility.
Treating the expected utility as a likelihood function, we can introduce
the likelihood ratio
\be
\label{118}
\Lbd(\pi_j) \equiv - \ln \; \frac{U(\pi_j)}{\sup_j U(\pi_j)}\; .
\ee
This likelihood ratio is non-negative, having minimum at zero. The expected
likelihood is given by
\be
\label{119}
\Lbd \equiv \sum_j p_0(\pi_j) \Lbd(\pi_j) \;  .
\ee

With entropy (\ref{116}), under condition (\ref{18}) and relation (\ref{119}),
the information functional is given by the expression
\be
\label{120}
I[p_0(\pi) ] = \sum_j p_0(\pi_j) \ln p_0(\pi_j) +
\lbd \left [ \sum_j p_0(\pi_j) - 1 \right ] +
\mu \left [ \sum_j p_0(\pi_j) \Lbd(\pi_j) - \Lbd \right ]  \;  ,
\ee
where $\lambda$ and $\mu$ are Lagrange multipliers. This functional is
minimized with respect to $p_0(\pi_j)$, when
\be
\label{121}
\frac{\dlt I[p_0( \pi)]}{\dlt p_0(\pi_j)} = 0 \; , \qquad
\frac{\dlt^2 I[p_0( \pi)]}{\dlt p_0(\pi_j)^2} > 0\; .
\ee
The corresponding variation derivatives yield
$$
\frac{\dlt I[p_0( \pi)]}{\dlt p_0(\pi_j)} = \ln p_0(\pi_j) + 1
+ \lbd + \mu \Lbd(\pi_j) \; ,
$$
\be
\label{122}
\frac{\dlt^2 I[p_0( \pi)]}{\dlt p_0(\pi_j)^2} = \frac{1}{p_0(\pi_j)}
> 0 \; .
\ee From the first of equations (\ref{121}), using (\ref{118}), and
denoting \be \label{123} Z \equiv e^{1+\lbd} \left [ \sup_j U(\pi_j)
\right ]^\mu \; , \ee which, under normalization (\ref{18}), becomes
\be \label{124} Z = \sum_j \left [ U(\pi_j) \right ]^\mu \; , \ee we
obtain the utility factor \be \label{125} p_0(\pi_j) = \frac{1}{Z}
\left [ U(\pi_j) \right ]^\mu \;  . \ee Note that, as long as
$\mu>0$, $ \left [ U(\pi_j) \right ]^\mu$ also defines a utility
function which results in the same ordering as the initial function
$U(\pi_j)$. Hence, expression (\ref{125}) gives an explicit relation
between the utility factor in QDT and the expected utility of
classical decision theory. From this relation, we see that QDT
reduces to classical decision theory when the interference terms
vanish so that $p_0(\pi_j) = p(\pi_j)$. The condition $\mu = 1$  in
expression (\ref{125}) is not necessary since $ \left [ U(\pi_j)
\right ]^\mu$ and $U(\pi_j)$ are two utility functions that result
in the same preference ordering. It is natural that the specific
value of the Lagrange multiplier $\mu$ should be irrelevant in the
correspondence between QDT to classical utility theory since $\mu$
just quantifies the degree to which  the likelihood can vary around
some a priori expected likelihood.

\section{Conclusions}

We have presented a quantum theory of decision making. By its nature,
it can, of course, be realized by a quantum object, say, by a quantum
computer or another quantum system. This theory provides a guide for
creating {\it thinking quantum systems} \cite{73}. It can be used as a
scheme for quantum information processing and for creating artificial
intelligence based on quantum laws. This, however, is not compulsory.
And the developed theory can also be applied to non-quantum objects with
an equal success. It just turns out that the language of quantum theory
is a very convenient tool for describing the process of decision making
performed by any decision maker, whether quantum or not. In this language,
it is straightforward to characterize entangled decisions,
non-commutativity of subsequent decisions, and intention interference.
These features, although being quantum in their description, at the same
time, have natural and transparent interpretations in the simple
everyday language and are applicable to the events of real life. To
stress the applicability of the approach to the decision making of
human beings, we have provided a number of simple illustrative examples.

We have demonstrated the applicability of the approach to the cases when
the Savage sure-thing principle is violated, resulting in the disjunction
effect. Interference of intentions, arising in decision making under
uncertainty, possesses specific features caused by aversion to uncertainty.
The theorem on interference alternation that we have derived connects the
aversion to uncertainty to the appearance of negative interference terms
suppressing the probability of actions. At the same time, the probability
of the decision maker not to act is enhanced by positive interference
terms. This alternating nature of the intention interference under
uncertainty explains the occurrence of the disjunction effect.

The theory has led naturally to a calculational method of the
interference terms, based on considerations using robust assessment
of probabilities, which makes it possible to predict their influence
in a {\it quantitative} way. The estimates are in good agreement
with experimental data for the disjunction effect.

The conjunction fallacy is also explained by the presence of the
interference terms. A series of experiments are analyzed and shown to
be in excellent agreement with the a priori evaluation of interference
effects. The conjunction fallacy is also shown to be a sufficient condition
for the disjunction effect, and novel experiments testing the combined
interplay between the two effects are suggested.

We have emphasized that the intention interference results in the
non-commutativity of subsequent decisions, which follows from the
theorem on non-commutativity of intended actions.

The approach of entropy maximization, or information-functional minimization,
is employed for deriving a relation between the quantum and classical
decision theories.

The specific features of the Quantum Decision Theory, distinguishing it
from other approaches known in the literature on decision making and
information processing, can be summarized as follows.

\vskip 2mm

(1) QDT is a general mathematical approach that is applicable to
arbitrary situations. We do not try to adjust the QDT to fit particular
cases; the same theory is used throughout the paper to treat quite
different effects.

\vskip 2mm

(2) Each decision maker is characterized by its own strategic state. This
strategic state of mind is, generally, not a trivial wave function, but
rather a composite vector, incorporating a great number of intended
competing actions.

\vskip 2mm

(3) QDT allows us to characterize not a single unusual, quantum-like,
property of the decision making process, but several of these
characteristics, including entangled decisions, non-commutative decisions,
and the interference between intentions.

\vskip 2mm

(4) The literature emphasizes that aversion with respect to uncertainty is
an important feeling regulating decision making. This general and ubiquitous
feeling is formulated under the uncertainty-aversion principle, connecting
it to the signs of the alternating interference terms.

\vskip 2mm

(5) The theorem on interference alternation is proved, which shows that
the interference between several intentions, arising under uncertainty,
consists of several terms alternating in sign, some being positive and
some being negative. These  terms are the source of the different
paradoxes and logical fallacies presented by humans making decisions in
uncertain contexts.

\vskip 2mm

(6) Uncertainty aversion and interference alternation, combined together,
are the key factors that suppress the probability of acting and, at the
same time, enhance the probability of remaining passive, in the case of
uncertainty.

\vskip 2mm

(7) The principal point is that it is not simply the interference between
intentions as such, but specifically the interference alternation,
together with the uncertainty aversion, which are responsible for the
violation of the Savage's sure-thing principle at the origin of the
disjunction effect.

\vskip 2mm

(8) The conjunction fallacy is another effect that is caused by the
interference of intentions, together with the uncertainty-aversion
principle. Without the latter, the conjunction effect cannot be
explained.

\vskip 2mm

(9) The conjunction fallacy is shown to be a sufficient condition
for the disjunction effect to occur, exhibiting a deep link between the
two effects.

\vskip 2mm

(10) The general ``interference-quarter law'' is formulated, which
provides a quantitative prediction for the amplitude of the
interference terms, and thus of the quantitative level by which the
sure-thing principle is violated.

\vskip 2mm

(11) Detailed quantitative comparisons with experiments documenting the
disjunction effect and the conjunction fallacy confirm the validity of
the derived laws.

\vskip 2mm

(12) Subsequent decisions are shown, in general, to be not
commutative with each other, by proving a theorem on
non-commutativity of decisions.

\vskip 2mm

(13) The minimization of an information functional, which is equivalent
to the conditional maximization of entropy, makes it possible to connect the
quantum probability with expected utility.

\vskip 2mm

(14) The relation between the quantum and classical decision theories
is established, showing that the latter is the limit of the former under
vanishing interference terms.

\newpage

{\large{\bf Acknowledgements}}

\vskip 3mm
We are very grateful to E.P. Yukalova for many discussions and useful advice
and to Y. Malevergne for stimulating feedbacks on an earlier version of the
manuscript. We acknowledge a helpful and illuminating correspondence with
P.A. Benioff and J.R. Busemeyer, which helped us to essentially improve the
presentation of the developed approach.

\end{document}